\documentclass[acmsmall,natbib=true]{acmart}
\usepackage{caption}
\usepackage{amsmath,amsfonts}
\usepackage{algorithmic}
\usepackage{algorithm}
\usepackage{array}
\usepackage[caption=false,font=normalsize,labelfont=sf,textfont=sf]{subfig}
\usepackage{textcomp}
\usepackage{stfloats}
\usepackage{url}
\usepackage{verbatim}
\usepackage{graphicx}
\usepackage{cite}
\usepackage{color}
\usepackage{listings}
\usepackage{xcolor}
\usepackage{booktabs}
\usepackage{multirow}
\usepackage{natbib}
\usepackage{tcolorbox}

\usepackage[commandnameprefix=always]{changes}
\usepackage{color}

\acmJournal{TAAS}
\makeatletter
\renewcommand{\paragraph}{\@startsection{paragraph}{4}{0pt}
    {3.25ex \@plus1ex \@minus.2ex}%
    {-1em}%
    {\normalfont\normalsize\bfseries}}
\makeatother

\title[DeFeed]{DeFeed: Secure Decentralized Cross-Contract Data Feed in Web 3.0 for Connected Autonomous Vehicles }

\thanks{This work is supported by the National Key R\&D Program of China (No.2023YFB2704100), National Natural Science Foundation of China (No.62472022, 62202038, 62272031, 62302022) and Beijing Nova Program (No.20230484257).}

\author{XINGCHEN SUN}
\affiliation{%
  \institution{Beijing Key Laboratory of Security and Privacy in Intelligent Transportation, Beijing Jiaotong University}
  \city{Beijing}
  \country{China}
}
\email{22120504@bjtu.edu.cn}
\author{RUNHUA XU}
\affiliation{%
  \institution{School of Computer Science and Engineering, Beihang University}
  \city{Beijing}
  \country{China}
}
\email{runhua@buaa.edu.cn}
\author{Wei Ni}
\affiliation{%
  \institution{School of Electrical and Data Engineering, University of Technology Sydney}
  \city{Sydney}
  \country{Australia}
}
\email{22110125@bjtu.edu.cn}
\author{LI DUAN}
\authornote{Corresponding authors: Chao Li and Li Duan}
\affiliation{%
  \institution{Beijing Key Laboratory of Security and Privacy in Intelligent Transportation, Beijing Jiaotong University}
  \city{Beijing}
  \country{China}
}
\email{duanli@bjtu.edu.cn}
\author{CHAO LI}
\authornotemark[1]
\affiliation{%
  \institution{Beijing Key Laboratory of Security and Privacy in Intelligent Transportation, Beijing Jiaotong University}
  \city{Beijing}
  \country{China}
}
\email{li.chao@bjtu.edu.cn}

\date{March 2024}

\begin{document}

\begin{abstract}
Smart contracts have been a topic of interest in blockchain research and are a key enabling technology for Connected Autonomous Vehicles (CAVs) in the era of Web 3.0. 
These contracts enable trustless interactions without the need for intermediaries, as they operate based on predefined rules encoded on the blockchain.
However, smart contacts face significant challenges in cross-contract communication and information sharing, making it difficult to establish seamless connectivity and collaboration among CAVs with Web 3.0.
In this paper, we propose \texttt{DeFeed}, a novel secure protocol that incorporates various gas-saving functions for CAVs, originated from in-depth research into the interaction among smart contracts for decentralized cross-contract data feed in Web 3.0.
\texttt{DeFeed} allows smart contracts to obtain information from other contracts efficiently in a single click, without complicated operations.
We judiciously design and complete various functions with \texttt{DeFeed}, including a pool function and a cache function for gas optimization, a subscribe function for facilitating data access, and an update function for the future iteration of our protocol.
Tailored for CAVs with Web 3.0 use cases, \texttt{DeFeed} enables efficient data feed between smart contracts underpinning decentralized applications and vehicle coordination.
Implemented and tested on the Ethereum official test network, \texttt{DeFeed} demonstrates significant improvements in contract interaction efficiency, reducing computational complexity and gas costs. Our solution represents a critical step towards seamless, decentralized communication in Web 3.0 ecosystems.
\end{abstract}




\maketitle

\renewcommand{\shortauthors}{Sun et al.}

\section{Introduction}
\label{sec:intro}


Connected Autonomous Vehicles (CAVs) have emerged as a transformative application domain that can benefit significantly from Web 3.0 technologies like blockchain \citep{nakamoto2008bitcoin, consensus-algorithm-review}, decentralized applications (dApps), and virtual or augmented reality. 
CAV networks rely on intelligent communication, coordination, and automated transactions \citep{feng2020magmonitor}.
These requirements are aligned with the vision of a more decentralized and trustless computing paradigm in Web 3.0.
A crucial factor for the development of Web 3.0 ecosystems is smart contracts, which are self-executing programs on the blockchain and allow for trusted interactions without intermediaries.
Smart contracts are particularly relevant for coordinating fleets of CAVs by encoding rules around negotiation, charging, access control, and other aspects of vehicle operation directly on the blockchain represented by the Ethereum.
The Ethereum \citep{buterin2014next, wood2014ethereum, wang2019blockchain} has the world's largest open-source community for blockchains, and smart contracts \citep{szabo1994smart,li2021silentdelivery} are its core technology. 
The combination of CAVs with Web 3.0 technologies like Ethereum's smart contracts has the potential to enable secure, decentralized coordination and automation of various vehicle functions. 
Smart contract-based dApps running on blockchain networks could help manage charging, ridesharing, vehicle access permissions, dynamic pricing models, and other services in a trustless manner without centralized authorities. 
Furthermore, blockchain's append-only ledger provides a tamper-proof record of vehicle operations, transactions, and interactions, which enhances transparency and accountability. 

In the context of CAV networks, the adaptivity and autonomy enabled by blockchain and smart contracts are particularly valuable.
Smart contracts can dynamically adjust operational parameters based on real-time data inputs, ensuring optimal performance without human intervention.
For example, smart contracts can automate the negotiation and execution of charging schedules for electric vehicles, adjusting based on current energy prices and network load.
This dynamic adaptation not only improves efficiency but also reduces operational costs.

Moreover, the decentralized nature of blockchain ensures that decisions within CAV networks are made transparently and without the need for a central authority \citep{yang2023zero}.
This enhances the reliability and trustworthiness if the system, as each transaction and decision is recorded on the blockchain and can be audited.
Such autonomous decision-making capabilities are crucial for the scalability and robustness of CAV networks, as they enable vehicles to operate independently while adhering to globally agreed-upon rules encoded in smart contracts.

Smart contracts fundamentally alter the way that Ethereum transactions occur \citep{li2020nf, li2020eventwarden, shi2023ress}, due to their many unique features, such as immutability, trustless execution, and high availability.
Once a smart contract is deployed, the code cannot be altered, ensuring that the original terms are preserved throughout the lifetime of the contract. 
The immutability prevents any party of the contract from unilaterally altering the agreement.
Smart contracts execute automatically based on their programmed logic and predefined conditions. 
They do not require a trusted third party or intermediary, allowing anonymous parties to conduct trustless transactions.
These contracts execute on-chain designed to be highly redundant and resistant to downtimes or single points of failure, ensuring high reliability.
With these desirable merits, smart contracts have gained widespread trust and play a crucial role in revolutionizing several rapidly expanding industries.
For instance, Decentralized Finance (DeFi) platforms, such as Uniswap \footnote{\url{https://uniswap.org/}} leverage smart contracts to provide better financial solutions.
Similarly, Non-Fungible Token (NFT) \citep{nft} trading facilitated by platforms like OpenSea \footnote{\url{https://opensea.io/}} and Rarible \footnote{\url{https://rarible.com/}}, have also witnessed significant growth due to the implementation of smart contracts.

\begin{figure*}[!t]\small
\centering
\includegraphics[width=5in]{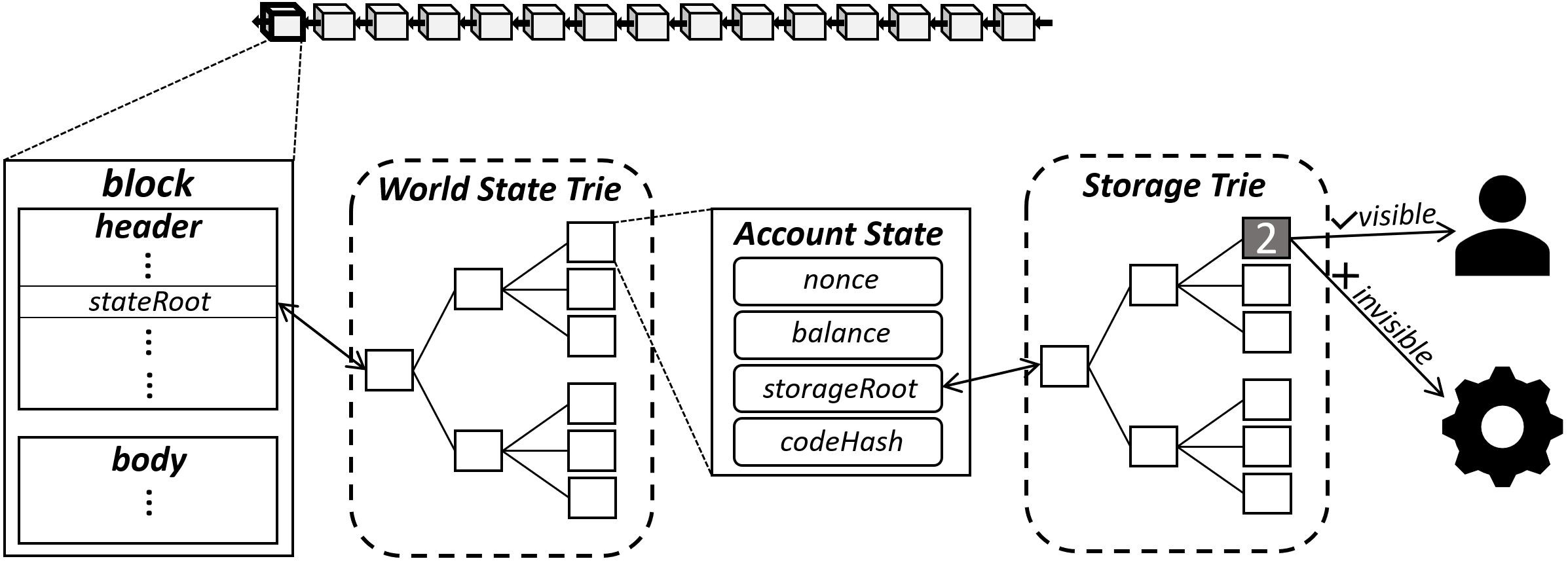}
\caption{ The  Cross-Contract  Data  Feed (CCDF) problem. The CCDF problem is represented in this structure. It arises because a smart contract cannot directly access variables stored in another contract's Storage Trie. For example, if an integer variable with a specific value is stored in one contract's storage, another contract doesn't have direct visibility to this data.}
\label{fig:problem}
\end{figure*}

Over the past decade, extensive research has focused on designing smart contracts to facilitate interactions between on-chain and off-chain systems. 
Various approaches have emerged to address data integration challenges.
The data feed mechanism \citep{signingtls} was initially developed to provide smart contracts with access to external world information. 
Early systems leveraged trusted hardware and blockchain technology to authenticate and ensure the reliability of external data sources. This innovation bridged the gap between smart contracts and trusted external websites, offering a foundation for reliable data authentication.
Subsequent advancements enhanced these capabilities significantly.
Improved data feed services introduced advanced features, such as enhanced transparency and consistency validation, enabling better parsing of external content and stronger authenticity guarantees. Later, privacy-preserving data feed schemes emerged, incorporating advanced cryptographic techniques like zero-knowledge proofs and privacy-focused technologies to secure sensitive data while maintaining its usability in smart contract operations. These developments represent a continuous evolution of data feed mechanisms toward greater security, reliability, and functionality.
Despite these advancements, existing approaches primarily focused on on-chain and off-chain interactions, overlooking a critical challenge: inter-contract data exchange. Within the Ethereum network, smart contracts remain largely isolated, unable to seamlessly share data with one another. This limitation significantly impedes potential collaborative and interconnected blockchain applications.
The Cross-Contract Data Feed (CCDF) problem represents a crucial research frontier, highlighting the need for innovative solutions to enable more dynamic and integrated smart contract ecosystems. Section \ref{subsec:problem} provides a comprehensive analysis of this challenge, as illustrated in Fig.~\ref{fig:problem}. 

In this paper, we propose \texttt{DeFeed}, an innovative secure protocol that incorporates various gas-saving functions for solving the decentralized CCDF problem among smart contracts.
The development of \texttt{DeFeed} is founded on our extensive research into the intricate interactions and communication mechanisms between different smart contracts deployed on the Ethereum.
One of the key challenges that \texttt{DeFeed} addresses is the inherent limitation in Ethereum's architecture that restricts direct data access across contracts.
This often requires complex operations and incurs substantial gas costs for smart contracts to obtain information from one another.
\texttt{DeFeed} overcomes the limitation by designing a new protocol architecture consisting of a series of interconnected smart contracts.
This empowers smart contracts with the ability to seamlessly obtain information from other contracts through a simple, one-click process.
As a result, \texttt{DeFeed} eliminates the need for complicated operations while maintaining low gas costs for cross-contract data feed.
Furthermore, we design and implement various gas-saving and efficiency-enhancing features within the \texttt{DeFeed} protocol. 

Notably, \texttt{DeFeed} incorporates a pool and cache mechanism to further optimize gas consumption.
The pool function aggregates multiple requests targeting the same data source, processing them in a batch to reduce redundant computations. 
The cache function pre-stores and caches frequently accessed data and function call results, eliminating the need for repetitive on-chain operations. 
These mechanisms significantly reduce the gas expenditure required for cross-contract data feed.
Additionally, \texttt{DeFeed} offers a subscribe function that enables smart contracts to subscribe to specific states from other contracts. 
Subscribed contracts automatically receive updates whenever the target contract's data changes, eliminating the need for repetitive querying. 
This subscription-based approach provides a more efficient and responsive data feed system, where contracts can stay up-to-date with the latest information without incurring the gas costs associated with constant polling.
Moreover, we integrate an update function into \texttt{DeFeed}.
As \texttt{DeFeed} is implemented as a series of smart contracts, it can undergo on-chain updates \footnote{\url{https://docs.openzeppelin.com/learn/upgrading-smart-contracts}}, facilitating the iterative improvement and evolution of its functionalities over time \citep{smartian, precise-attack-synthesis, ethbmc, elysium, grech2018madmax, contractfuzzer, li2024game, bumiller2023understanding, luu2016making, sfuzz, xue2022xfuzz, multimediaWorkflow}.

By leveraging the pool, cache, subscribe, and update features, \texttt{DeFeed} significantly improves the cost-effectiveness and responsiveness of the cross-contract data feed process on the Ethereum blockchain, making it a more practical and scalable solution for smart contract interactions.
With the \texttt{DeFeed} protocol, smart contract developers can effortlessly access cross-contract data through a simplified interface, without the burden of intricate operations. 
This not only enhances the development experience but also potentially unlocks new possibilities for decentralized applications by enabling efficient and secure data feed across the Ethereum ecosystem.

\noindent \textbf{Contributions.}The contributions of this paper are summarized as follows:
\begin{itemize}
\item To the best of our knowledge, \texttt{DeFeed} is the first secure and decentralized cross-contract data feed protocol designed for Ethereum-like platforms.
\texttt{DeFeed} empowers smart contracts to access and exchange data from each other without compromising the core principles of decentralization, making it a pioneering protocol for CAVs with Web 3.0.
\item We utilize the state-of-the-art technology to develop and implement upgradable smart contracts with various functions, such as function pool and cache for gas-saving and efficiency, and a subscribe function to provide low-cost access to information in \texttt{DeFeed}.
\item We successfully implement and evaluate \texttt{DeFeed} on the Ethereum official test network. 
The results demonstrate that \texttt{DeFeed} is inexpensive and convenient to implement, and efficient for facilitating communication and data feed between smart contracts - a key enabler for Connected Autonomous Vehicles networks with Web 3.0.
\end{itemize}

The rest of this paper is organized as follows: we introduce the background and related work in Section \ref{sec:background} and provide an overview of the cross-contract data feed problem in Section \ref{sec:overview}.
In Section \ref{sec:defeed}, we propose some components and functions about \texttt{DeFeed}.
Section \ref{sec:imp} describes the performance evaluation of proposed \texttt{DeFeed}.
Finally, we conclude the paper in Section \ref{sec:conclusion}.

\section{Background and Related Work}
\label{sec:background}
In this section, we explore the link between autonomous and adaptive systems and blockchain, outline the basics of various Ethereum blockchain accounts, introduce essential cryptographic tools used in our work, and discuss related work.

\subsection{Autonomous and Adaptive Systems with Blockchain}

Autonomous and adaptive systems \citep{liu2024consortium} are designed to perform tasks without human intervention and to adjust their behavior in response to changes in their environment. These systems play a crucial role in various modern technologies, including CAVs, smart grids, and Internet of Things (IoT) networks \citep{su2023hybrid}. The primary aim of these systems is to enhance efficiency, reliability, and scalability while minimizing the need for centralized control and human oversight.

Autonomous systems operate independently, making decisions based on predefined rules and real-time data inputs. They are capable of performing tasks such as navigation, coordination, and resource allocation without direct human control. Key characteristics of autonomous systems include self-governance, real-time operation, scalability, and reliability. Self-governance refers to the system's ability to make decisions and perform actions based on internal logic and external stimuli. Real-time operation ensures that the system processes and responds to data in a timely manner, facilitating accurate and prompt actions. Scalability allows the system to handle increasing amounts of work or to expand easily to accommodate growth. Reliability guarantees consistent and predictable performance under various conditions.

Adaptive systems, on the other hand, are engineered to adjust their behavior based on changes in their operating environment. This adaptivity ensures that the system remains efficient and effective even as conditions evolve. Essential features of adaptive systems include dynamic adjustments, learning and optimization, and flexibility. Dynamic adjustments enable the system to modify operational parameters and strategies in response to new data or environmental changes. Learning and optimization allow the system to improve performance over time through learning algorithms and optimization techniques. Flexibility provides the system with the ability to handle a wide range of scenarios and conditions without requiring significant reconfiguration or intervention.

Blockchain technology \citep{jiang2018blockchainiov} and smart contracts are highly relevant to the development and enhancement of autonomous and adaptive systems. Blockchain offers a decentralized, secure, and transparent platform for data storage and transaction processing, which is critical for the trustless operation of these systems \citep{wang2023fastv2v}. Smart contracts, which are self-executing programs running on the blockchain, facilitate automation and dynamic interactions among system components.

In the context of CAV networks \citep{jabbar2021V2X}, blockchain and smart contracts offer several advantages. The decentralized nature of blockchain eliminates the need for a central authority, thereby reducing single points of failure and enhancing system resilience. 
Blockchain's immutable ledger ensures that all transactions and data are secure and verifiable, fostering trust among participants. 
Smart contracts automate various processes, such as vehicle coordination, charging, and access control, based on predefined rules and real-time data inputs. 
Furthermore, smart contracts can dynamically adjust their operations in response to changes in the environment, such as fluctuating energy prices or varying traffic conditions, thereby enhancing the system's adaptivity.

\subsection{Preliminaries for Ethereum}

\noindent \textbf{External Owned Accounts:}
External Owned Accounts (EOAs) are Ethereum accounts that are controlled by private keys outside the Ethereum network. 
This is the most common type of account on the Ethereum blockchain, and individuals commonly use it to interact with the Ethereum network, store Ether and other ERC-20 tokens, and send or receive transactions.

EOAs are identified by their Ethereum address, a 20-byte hexadecimal string. 
When an EOA initiates a transaction on the Ethereum network, it must sign the transaction with its private key and broadcast it to the network. 
Once the transaction is confirmed, the Ethereum network updates the state of the account accordingly.
EOAs differ from Contract Accounts (CAs), which are accounts associated with smart contracts deployed on the Ethereum network. 
CAs are controlled by the smart contract's code, rather than by a private key, and are used to store and manage assets or execute complex transactions.

In summary, EOAs are a fundamental and essential component of the Ethereum network as they allow for secure and decentralized interaction between individuals and the network.

\noindent \textbf{Transactions:}
Ethereum supports two main types of transactions: regular value transfers and contract interactions.
Regular transactions involve the simple transfer of Ether from one account to another.
On the other hand, contract interactions require the invocation of functions or execution of code within smart contracts deployed on the Ethereum blockchain \citep{li2024twatch}.

Each transaction in Ethereum comes with a unique identifier called a nonce that prevents double-spending attacks and ensures that transactions from a specific account are executed in the correct order \citep{li2023howhard}.
The nonce increments with each transaction sent from an account.
When it comes to executing transactions in Ethereum, the gas limit is a crucial factor to consider.
It prevents the maximum amount of gas that a sender is willing to spend on a transaction, which determines the computational resources, such as CPU time and memory, allocated to execute the transaction.
If the gas limit is set too low, the transaction may run out of gas and fail.
Hence, setting an appropriate gas limit is crucial to ensure that transactions execute successfully. 

Another essential factor to consider is the gas price. 
It refers to the amount of ether that the sender is willing to pay per unit of gas used in the transaction. 
Miners prioritize transactions with higher gas prices, as they provide greater incentives for including them in blocks. 
Moreover, gas prices are not static and can fluctuate depending on the level of network traffic and the prevailing demand in the market.

Transaction fees in Ethereum are calculated as 
\begin{equation}
TotalFee = (baseFee + priorityFee) * gasPrice 
\end{equation}

The total fee is determined by multiplying the gas price by the actual gas consumed during transaction execution.
Gas consumption depends on the complexity of the transaction or smart contract operation.
When a transaction is initiated, it is broadcast to the Ethereum network and propagated to nodes. Miners then include pending transactions in blocks by solving complex mathematical puzzles through the mining process. 
The blockchain consensus mechanism guarantees that a valid transaction takes at most $\theta$ time to be recorded on the ledger \citep{nakamoto2008bitcoin, xu2022reputation, xu2021concurrent}. 
The time delay $\theta$ is also known as the blockchain delay.
Once a block is mined and added to the blockchain (e.g., the ledger), the transaction is considered confirmed.
While transactions are considered final once they are confirmed and added to the blockchain, it's essential to note that Ethereum employs a probabilistic finality mechanism. 
This means that as more blocks are added to the blockchain after a transaction's inclusion, the probability of its reversal decreases significantly.
It can only be changed through a network-level attack that costs billions of dollars.

\noindent \textbf{Ethereum Virtual Machine:}
The Ethereum Virtual Machine (EVM) is a key component of the Ethereum blockchain that enables the execution of smart contracts and decentralized applications (DApps). 
It serves as a decentralized runtime environment for executing code written in Ethereum's native programming language, Solidity \footnote{\url{https://github.com/ethereum/solidity}}, as well as other compatible languages like Vyper \footnote{\url{https://github.com/vyperlang/vyper}}.

The EVM is a decentralized execution environment that runs on thousands of nodes in the Ethereum network. 
Each node keeps a copy of the Ethereum blockchain and executes smart contracts and transactions independently. 
Smart contracts are deployed on the Ethereum blockchain and are converted into bytecode, which is a low-level representation of the contract's instructions. 
The EVM then executes this bytecode sequentially, interpreting each operation code and performing the corresponding action. 
These operation code sequences are generated from the high-level smart contract code written in Solidity or other compatible languages.
The EVM guarantees deterministic execution of smart contracts, which means that when given the same input and initial state, the outcome of contract execution is always the same. 
This property is crucial for achieving consensus among nodes on the network, as all nodes must arrive at the same result when executing transactions and smart contracts.

\noindent \textbf{GAS:}
On the Ethereum network, users pay Ethereum gas \citep{technicalSpecification} to process transactions or use smart contracts. 
Ethereum gas is measured in gwei, which is short for gigawei, with one gwei being equivalent to one billionth of an ETH. 
However, Ethereum gas fees can only be paid using Ether, which is the platform's native token.

It's crucial to note that a malicious user posing as a legitimate one can execute resource-exhausted code in a function, leading to a Denial of Service (DoS) attack that could harm the entire network.
This is because every single node in the network has to execute every function call to achieve a consensus on the state of balances and contracts.

To safeguard against such attacks, each fundamental atomic operation has been assigned a specific amount of gas.
Each Ethereum transaction has a specified gas limit, which represents the maximum amount of gas the sender is willing to pay for the transaction. 
Gas price is the amount of Ether paid per unit of gas and is set by the sender. 
The total transaction fee is the product of the gas limit and gas price.
Hence, whenever a function is called, the initiator of the call must pay a transaction fee that covers the total gas consumption of the on-chain computations involved in the call. 
As an incentive, the fee above is paid to the miner, not to each individual node.

While gas has been effective in preventing DoS attacks, it has also led to a significant increase in transaction fees for users of blockchain \citep{Under-optimizedSmartContractsDevourYourMoney}. 
According to the statistics of the Ethereum blockchain, the largest programmable blockchain in terms of market capitalization \footnote{\url{https://coinmarketcap.com/}}, users had to pay an average of 5,369,200 ETH (approximately 10 billion USD) per day in gas costs in the last year.
The official documentation of the Ethereum Foundation has acknowledged the issue of high gas fees \footnote{\url{https://ethereum.org/en/developers/docs/gas/}}.

Due to the high fees, several smart contracts have limited functionality and rely on simple programs where each function ends in a constant number of steps. 
Additionally, there are several protocols, including layer-two protocols, that sacrifice decentralization or trustlessness in order to reduce gas costs.

\subsection{Cryptographic tools}
\noindent \textbf{Keccak-256 hash function:}
The Keccak-256 hash function belongs to the Keccak family of cryptographic hash functions. It gained widespread recognition for being chosen as the algorithm for Secure Hash Algorithm 3 (SHA-3) by the National Institute of Standards and Technology (NIST) in 2012. However, it's worth noting that Ethereum's use of Keccak-256 precedes the final standardization of SHA-3, and as a result, it differs slightly from the SHA-3-256 variant standardized by NIST.

In Ethereum, Keccak-256 is used extensively for various purposes.
Creating Addresses, Ethereum addresses are derived by taking the Keccak-256 hash of the public key derived from the private key and then taking the last 40 characters (20 bytes) of that hash.
Transaction Hashing, transactions are identified by their Keccak-256 hash.
Smart Contracts, the bytecode of Ethereum smart contracts is also hashed using Keccak-256. This hashing facilitates the EVM operation and integrity verification.
Merkle Patricia Trees (MPT), Ethereum uses a modified version of a Merkle Patricia Tree for its blockchain state, where Keccak-256 is used to hash the nodes and leaves of the tree, ensuring integrity and enabling efficient and secure data verification.

\noindent \textbf{ECDSA:}
Many mainstream blockchain systems, such as Ethereum and Bitcoin, adopt the Elliptic Curve Digital Signature Algorithm(i.e., ECDSA). 
ECDSA is a public key cryptography algorithm used to verify the validity of digital signatures. 
It is based on elliptic curve cryptography and uses private and public keys to generate digital signatures and verify their validity. 
In Ethereum, ECDSA is used to verify whether the sender of a transaction has the authority to send it. 
Specifically, when a user sends a transaction, they must sign it with their private key and then broadcast the signature and transaction together on the network. 
Other nodes can use the sender's public key to verify the validity of the signature, ensuring that the transaction was sent by its rightful owner. 
In Ethereum, ECDSA is also used to generate Ethereum addresses. An Ethereum address is generated from a public key using hash functions, so ECDSA algorithms are needed for generating public keys. 
Typically, an Ethereum address consists of 40 hexadecimal characters in string format, such as 0x7cB57B5A97eAbe94205C07890BE4c1aD31E4XXXX.

\subsection{Upgradeable Smart Contracts}
Immutability or tamper-proof is a major property of blockchain systems, like Ethereum.
Smart contracts that are deployed on such blockchain systems also possess this property.
Smart contracts play an important role in the operation of blockchain systems with this advantage.

But they also have certain flaws and limitations.
The primary flaw is that they can not be upgraded as general software does.
Once deployed on the blockchain systems, these contracts can not be altered or upgraded.
This means that any bugs, vulnerabilities, or shortcomings discovered after deployment cannot be easily rectified without deploying an entirely new contract.
Bugs or errors discovered in the smart contract code after deployment cannot be easily fixed in non-upgradeable contracts.
Users and developers may need to resort to deploying an entirely new contract, which can be a cumbersome and costly process.
If there is a need to migrate to a new contract (e.g., due to security concerns or major updates), users may face challenges in moving their assets and data from the old contract to the new one. 
This process can be complex and may result in fragmentation of the user base.
In the event of a security vulnerability, non-upgradeable contracts may pose significant risks. 
Without the ability to patch or upgrade the contract, users' funds and assets may be at risk, and exploits could lead to irreversible consequences.
How to design upgradeable smart contracts is a key issue of concern in both academic and industry circles.

The proxy pattern \citep{proxyhunting, safeCreationAndUpgradeOfEthereumSmartContracts} is a great design that smart contracts can use to be upgradeable.
It gives developers some kind of leeway to modify contract logic post-deployment.
In a proxy-contract architecture, there are two main parts:
\begin{enumerate}
\item [1)] The proxy contract,
\item [2)] The execution or logic contract.
\end{enumerate}
The proxy contract is deployed at the beginning and contains the contract storage and balance. 
The execution contract stores the contract logic that is used to execute functions.
The proxy contract stores the address of the execution contract.
When users send requests, the message goes through the proxy contract.
The proxy contract then routes the message to the execution contract, which performs the computation. 
Afterward, the proxy contract receives the result of the computation from the execution contract and returns it to the user.
It is important to note that the proxy contract itself cannot be modified. 
However, additional execution contracts with updated contract logic can be created, and message calls can be rerouted to the new contract. 
With the proxy pattern, you are not changing the smart contract itself. 
Instead, you are deploying a new logic contract and asking the proxy contract to reference it instead of the old contract. 
This new logic contract can have different functionality than the previous one or fix an old bug.

\subsection{Related Work}
In recent decades, blockchain technology has gained widespread adoption across various industries. One of its key features is the implementation of smart contracts, which enable the execution of business logic in a decentralized structure, resulting in trusted and universally accepted outcomes among participating nodes. However, smart contracts face a fundamental limitation: they cannot independently access external data sources. To address this, they rely on oracles - off-chain external data sources - to collect and provide data feeds and inputs \citep{town-crier}, \citep{trustworthy}. 

Extensive research has been conducted on blockchain oracles and data feed mechanisms. Zhang et al. \citep{town-crier} introduced TownCrier, a platform leveraging hardware-based Trusted Execution Environments (TEEs), such as Intel's SGX enclaves, to collect data from HTTP-enabled sources and provide authenticity proofs securely. This approach significantly enhanced the trustworthiness of external data inputs to smart contracts. 
Building on this foundation, Juan et al. \citep{pdfs} proposed the Practical Data Feed Service (PDFS) system, which establishes a robust connection between smart contracts and external data sources. PDFS addresses scalability and efficiency concerns in oracle systems, offering a more practical solution for real-world applications.

Recognizing the importance of privacy in blockchain data feeds, Wang et al. \citep{privacy-data-feed} developed an innovative on-and-off-chain data feed privacy-preserving scheme. This approach incorporates zero-knowledge proofs \citep{zero-knowledge-proof} and Hawk technology \citep{hawk}, enabling smart contracts to interact with external data while maintaining strong privacy guarantees.
The reliability and trustworthiness of oracle systems have been a focal point of research. Lo et al. \citep{reliability} conducted a comprehensive analysis of trust-enabling features across various blockchain oracle approaches, techniques, and platforms. Their work provides valuable insights into the strengths and weaknesses of different oracle solutions, guiding future developments in the field.

Recent developments in the field include the exploration of multi-chain oracle solutions to enhance interoperability between different blockchain networks. Additionally, researchers are investigating the use of artificial intelligence and machine learning techniques to improve the accuracy and reliability of data feeds.

Blockchain technology has been widely recognized for its potential to enhance the functionality of autonomous systems by providing a decentralized and secure platform for communication and decision-making. For instance, Menha et al. \citep{megha2020survey} explored the use of blockchain in autonomous vehicles, highlighting how blockchain can ensure secure data sharing and trust among vehicles.
Ren et al. \citep{ren2021double} addresses the security of IoT devices' sampled data in agriculture, proposing a double-blockchain solution using IPFS storage.
Rathee et al. \citep{rathee2019blockchainframework} employ blockchain technology to solve
the various security challenges in connected vehicles.

Smart contracts, which are self-executing programs on the blockchain, play a crucial role in enabling adaptivity in various systems. 
Baza et al. \citep{baza2019blockchainfirmware} employed the blockchain and smart contract that proposes a distributed firmware update scheme for autonomous vehicles.
Jain et al. \citep{jain2021blockchainAVs} discussed how smart contracts could automate processes in supply chain management, adapting to new data inputs and conditions without human intervention. In the context of Internet of Vehicles (IoV), Sigh et al. \citep{singh2020IoV} introduces a blockchain-based decentralized trust management scheme for IoV using smart contracts to address security and privacy threats.

To sum up, current techniques and tools for data feed in blockchains focus on enabling data interaction between on-chain and off-chain sources.
Our work in this paper focuses on the data feed between smart contracts on-chain and tackles the decentralized CCDF problem.
To the best of our knowledge, \texttt{DeFeed} is the first secure and decentralized data feed protocol designed for Ethereum-like platforms and CAVs with Web 3.0.

\section{The Cross-Contract Data Feed problem}
\label{sec:overview}
In this section, we provide an overview of the CCDF problem. We then introduce two existing architectures for solving this problem and propose a new solution employed by \texttt{DeFeed}. 
 
\subsection{Overview}
\label{subsec:problem}
The CCDF problem is essentially induced by the lack of visibility of a smart contract to variables stored in another smart contract.
In Fig.~\ref{fig:problem}, we illustrate where an integer variable with the value is stored in Ethereum.
When a new block is created, each miner must synchronize with the most recent block and update the local database storage (i.e., LevelDB in Ethereum) accordingly.
A block in Ethereum consists of a header and a body, with the header containing several elements, including the roots of three MPT trees, namely the root \textit{stateRoot} for the World State Trie tree and two other roots for the Receipts Trie tree and the Transactions Trie tree, respectively.
In the World State Trie tree, each leaf node represents a snapshot of the current state of an EOA or a CA.
The state of an account consists of four components:
\begin{equation}
  \label{e1}
  \textit{state}\ \equiv\ \langle 
  \textit{nonce}, \textit{balance}, \textit{storageRoot}, \textit{codeHash} \rangle
  \end{equation}
where \textit{nonce} indicates the number of transactions sent from the account, \textit{balance} denotes the current account balance, and \textit{codeHash} represents the hash value of the EVM bytecode of the smart contract in case the account is a CA.
Here, \textit{storageRoot} refers to the root of the Account Storage Trie tree for a CA, where the current values of all smart contract variables are stored.

The state of variables in smart contracts can be considered visible to miners because encoded values of these variables are physically stored in LevelDB, and miners can either use the verified contract code provided by platforms, such as Etherscan, or reverse engineering tools to decode and read the values.
In addition, public variables have built-in getters, so EOAs can always access them.
In contrast, private/internal variables might not be directly accessible to EOAs, although they could be made accessible to EOAs through explicit getters.

Miners and EOAs are essentially two roles people play, so they can read values of variables from smart contracts using various off-chain techniques as long as the blockchain data is available.
Unlike miners and EOAs, smart contracts, or CAs, are like the residents living in the blockchain world; hence, they have to obey the rules made by the developers of the blockchain system. 
Unfortunately, in Ethereum, one critical rule is that a smart contract is unaware of the state of variables in another smart contract.
In other words, by default, the storage of a smart contract is invisible to another smart contract, resulting in the CCDF problem.

\begin{figure}[!t]\small
\centering
\includegraphics[width=4.5in]{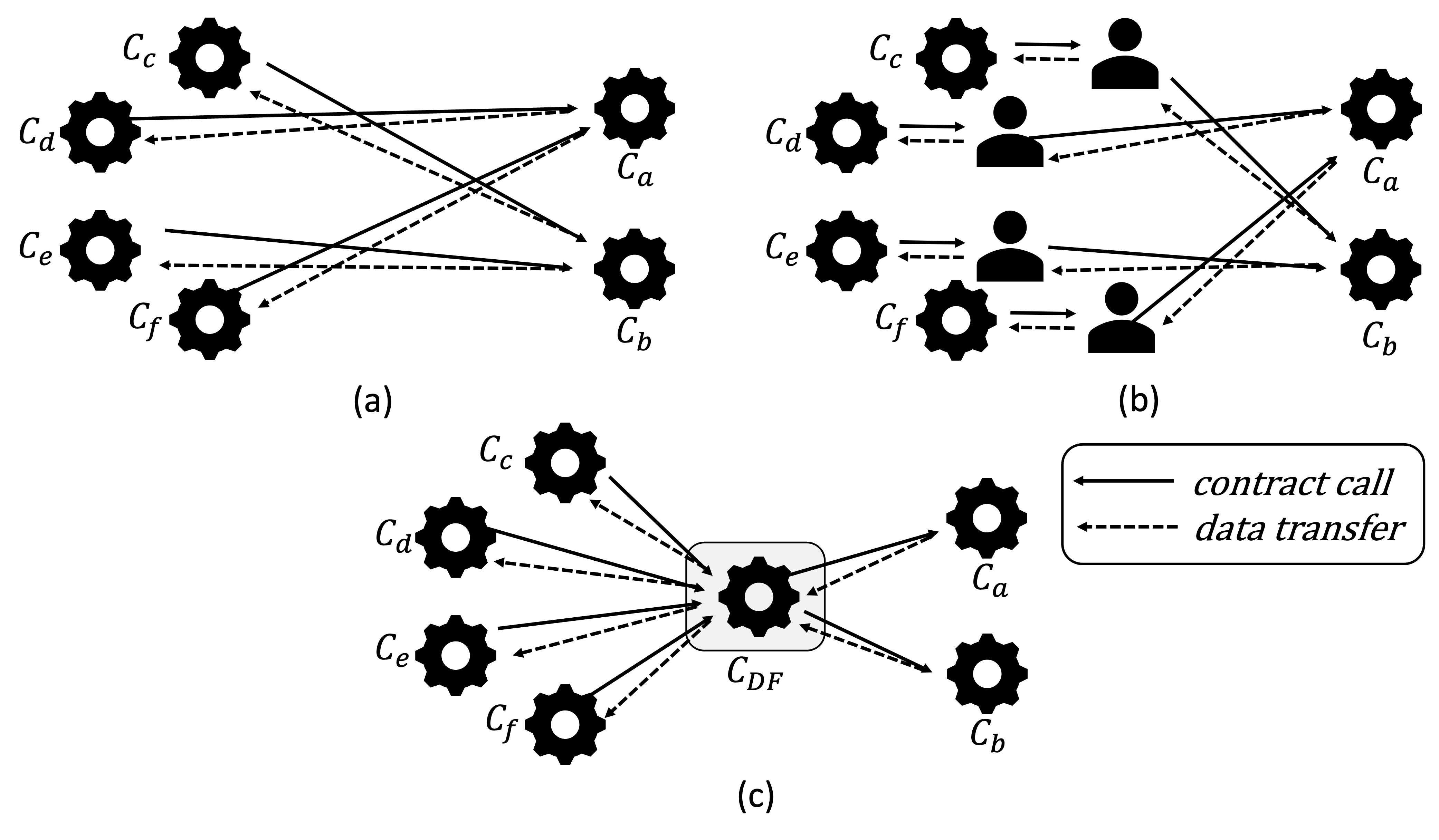}
\caption{The traditional data feed architecture (a), the third-party data feed architecture (b), and the cross-contract data feed architecture (c). In this figure, each gear represents a smart contract, each persona represents a third party. Solid arrows indicate "contract call". while dashed arrows represent "data transfer".}
\label{fig:architecture}
\end{figure}

\subsection{Architectures for CCDF}
In Ethereum, the traditional data interaction process can be implemented through different architectural approaches, as illustrated in Fig.~\ref{fig:architecture}.(a) and Fig.~\ref{fig:architecture}.(b). 
Here, $C_c$, $C_d$, $C_e$ and $C_f$ represent smart contracts that own specific data, while $C_a$ and $C_b$  represent smart contracts seeking to request data.

The direct data feed architecture, shown in Fig.~\ref{fig:architecture}.(a), establishes a direct connection between contracts like $C_a$ and $C_d$.
This architecture enables immediate data exchange between contracts while maintaining low gas costs. However, this approach raises significant privacy concerns due to its lack of access control mechanisms.

Fig.~\ref{fig:architecture}.(b) presents an alternative architecture that introduces a third party to mediate data feed between smart contracts. While this approach enables monitoring of data legality, it introduces additional costs and creates a single point of trust. Furthermore, the reliance on a centralized third party may lead to availability issues, potentially causing delays in data delivery.

To address these limitations, we propose a novel architecture illustrated in Fig.~\ref{fig:architecture}(c). This design replaces the third party with a dedicated smart contract ($C_{DF}$) that serves as a secure data feed hub. Data-providing contracts can register with $C_{DF}$ using contract names or specific nicknames, creating a binding between these identifiers and their addresses. 
When contract $C_a$ needs data from $C_d$, it sends a request to $C_{DF}$ containing the registered name of the target contact.
$C_{DF}$ then resolves the address and facilitates the data exchange, maintaining isolation between data owners and requestors while ensuring secure and efficient data transfer.

\subsection{The chainedCALL pattern}
To implement the proposed architecture, we employ a fundamental design pattern based on the chained opcode \texttt{CALL}. This pattern, which we term \textit{chainedCALL}, enables controlled data access between smart contracts through explicit function provisioning.

\begin{figure*}[!t]\small
\centering
\includegraphics[width=5.5in]{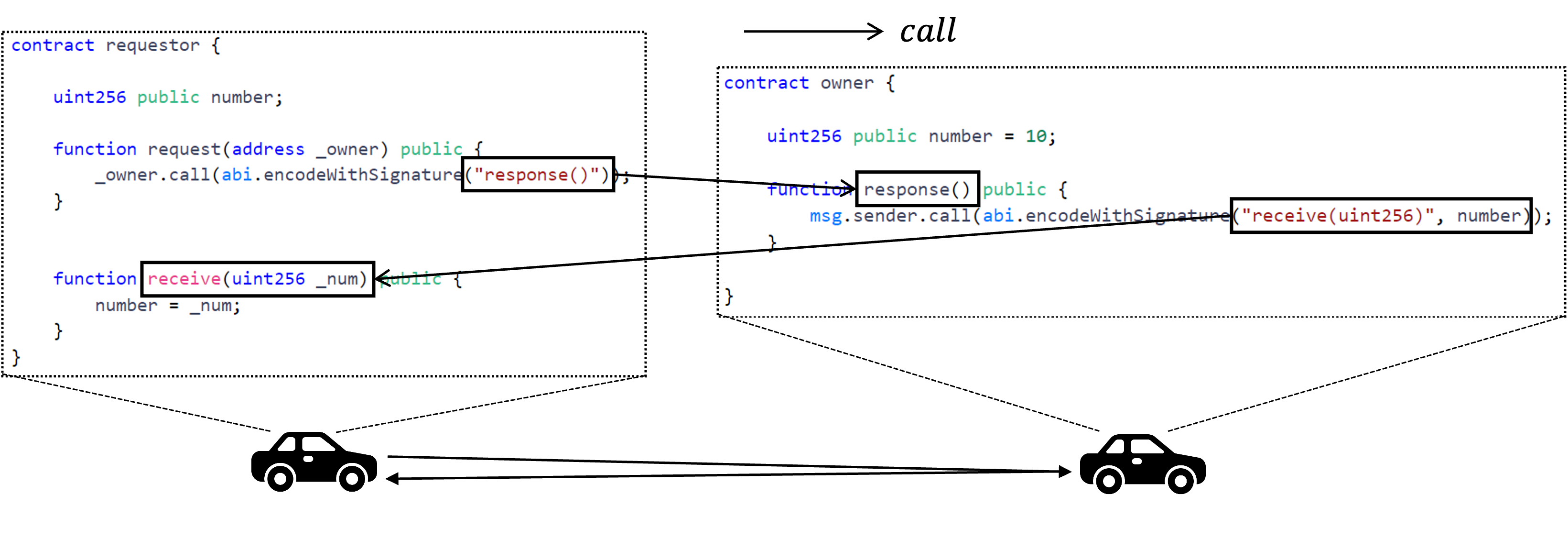}
\caption{Regular data feed process. Solid arrows represent contract function calls. The request function in the requester contract calls the response function in the owner contract, which then calls the receive function in the requester contract.}
\label{fig:chainedcall}
\end{figure*}

Fig.~\ref{fig:chainedcall} demonstrates the elementary implementation of this pattern through three dedicated functions: \textit{request}, \textit{response}, and \textit{receive}. The process begins when an EOA initiates a function call to \textit{request} in the data requestor contract. This triggers a chain of \texttt{CALL} operations, first invoking \textit{response(·)} in the data owner contract, which then calls \textit{receive(·)} in the requestor contract to complete the data transfer.

The chainedCALL pattern serves as the basic architectural building block for CCDF, represented by turn-around lines with arrows between contracts as shown in Fig.~\ref{fig:chainedcall}. While this example presents a simplified implementation, real-world applications require additional logic for error handling, data validation, and security measures to ensure reliable operation of CAVs and their interactions.

By combining our proposed architecture with the chainedCALL pattern, \texttt{DeFeed} achieves a balance between security, efficiency, and practical implementation requirements for cross-contract data feed operations.

In the rest of this paper, we consider the above pattern of the chained opcode \texttt{CALL} as the basic architecture building block for CCDF.
For simplicity, we name the pattern \textit{chainedCALL} and denote it by turn-around lines with arrows between two smart contracts, as shown in Fig.~\ref{fig:chainedcall}.

\section{The proposed method: DEFEED}
\label{sec:defeed}

In this section, we first present the basic protocol called \texttt{Data Feed}, which enables data feed across different smart contracts. 
Next, we present more complex protocols involving the function \textit{Pool} and \textit{Cache} based on the basic protocol.
Additionally, we explain how the protocol upgrades core components using the \textit{Update} function and how smart contracts can receive notifications when the subscribed smart contract updates its status with the function \textit{Subscribe}.

\subsection{Data feed}
\label{subsec:datafeed}

\begin{figure}[!t]\small
\centering
\includegraphics[width=3.5in]{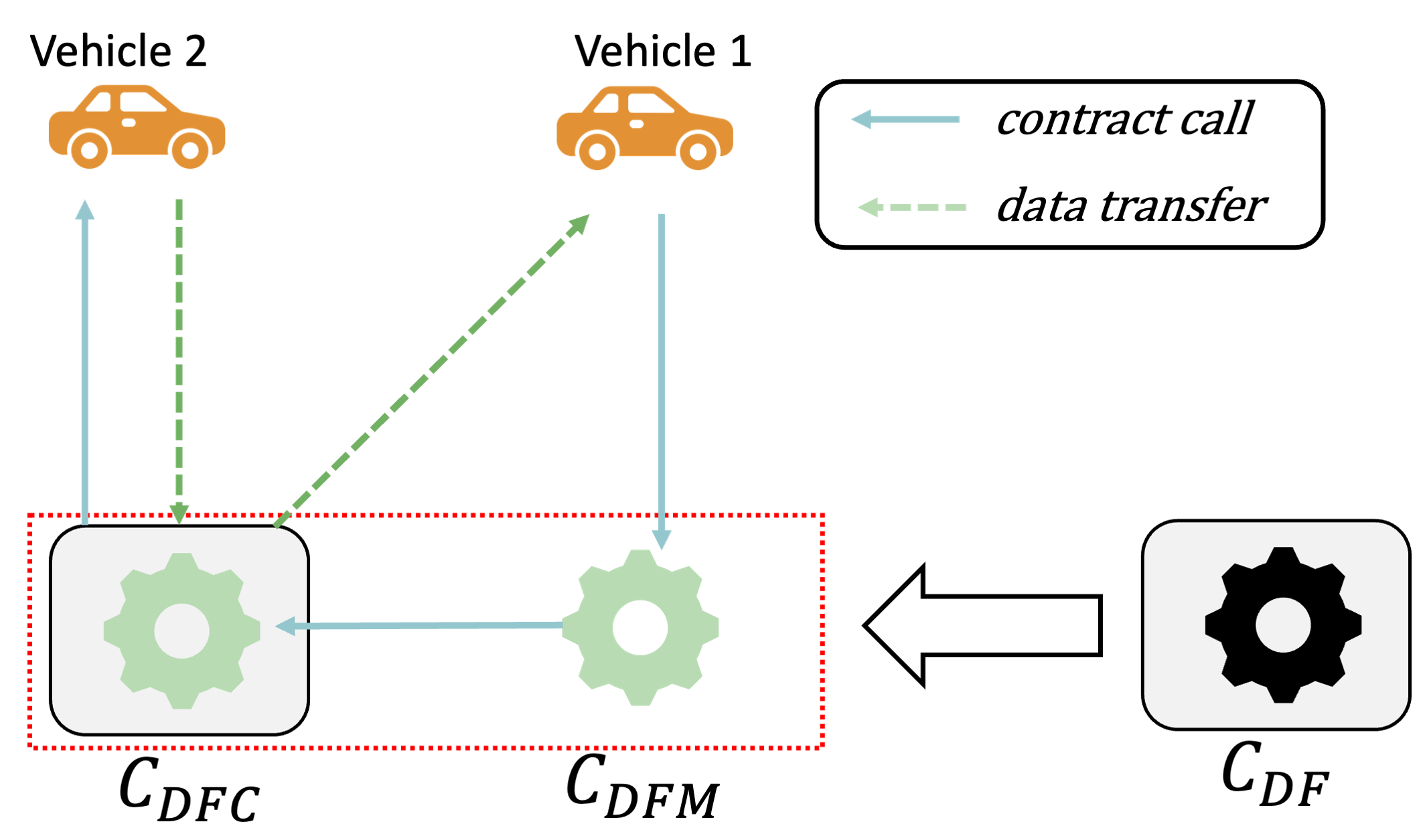}
\caption{Overall framework of DeFeed. The smart contract $C_{DF}$ is divided into two separate contracts. Solid blue arrows represent "contract call", while the dashed green arrows indicate "data transfer". Vehicle 1 requests data from Vehicle 2 using the complex contracts outlined in the red dashed box. A red box indicates that its interior is isolated from the exterior.}
\label{fig:overall}
\end{figure}

\begin{figure}[!t]\small
\centering
\includegraphics[width=3.5in]{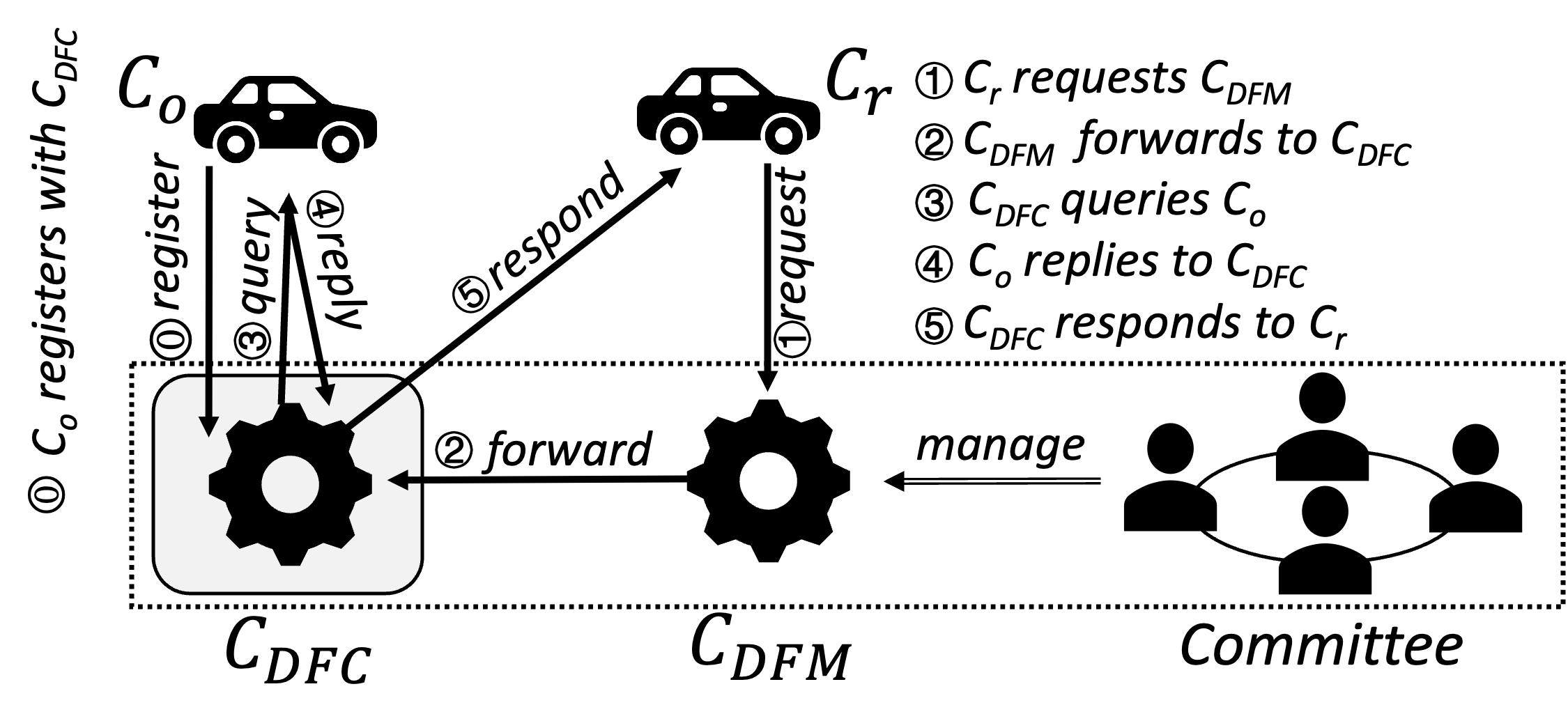}
\caption{Detailed workflow of the regular data feed process. 
A contract $C_{DFC}$ receives registration data from a vehicle $C_o$ (Step 0). Another vehicle $C_r$ can make requests to a contract $C_{DFM}$ contract for accessing the registered data (Step 1). $C_{DFM}$ will forward the request to $C_{DFC}$ (Step 2). $C_{DFC}$ queries $C_o$ for the data (Step 3\&4) and responds it to $C_r$ (Step 5).
The committee manages the overall process.}
\label{fig:regular}
\end{figure}

To establish a secure communication protocol among smart contracts, we first devise and implement a fundamental protocol called the \texttt{Data Feed}. 
To improve isolated management and facilitate future protocol iterations, we modified the architecture from Fig.~\ref{fig:architecture}(c) to the overall framework depicted in Fig.~\ref{fig:overall} for \texttt{DeFeed}.
We split one smart contract $C_{DF}$ into two separate smart contracts and create a decentralized committee in order to avoid the centralization of our protocol.
One of the smart contracts is responsible for management, while the other is responsible for implementing the basic functions of the protocol.
We sketch the protocol in Fig.~\ref{fig:regular} and present a formal description of data feed interactions.

The fundamental protocol comprises five primary components: a Data Owner Contract ($C_o$), a Requestor Contract ($C_r$), a Data Feed Center Contract ($C_{DFC}$), a Data Feed Management Contract ($C_{DFM}$), and a Committee.
The Committee is a collective of individuals responsible for managing $C_{DFM}$. 
The contract $C_o$ represents a CAV that provides data, while the contract $C_r$ represents a CAV that requests data.
The smart contracts $C_{DFC}$ and $C_{DFM}$ are the core part of \texttt{DeFeed}, facilitating data feed between parties.
We formally define the workflow system as a triple:
\begin{equation}
    W = (C, I, M)
\end{equation}
where $C$ is the set of components $C = \{C_o, C_r, C_{DFC}, C_{DFM}\}$, $I$ represents the interactions between these components, and MM denotes the management rules applied by the Committee.
Each interaction in the set $I$ signifies an action, such as a request initiation by $C_r$, a request forward by $C_{DFM}$, or a response by $C_{DFC}$.
The Committee's management role is crucial as it shapes the behavior and decision-making process within $C_{DFM}$, thereby influencing the overall efficiency of the protocol.
Next, we will provide a detailed description of the protocol.

\indent \textit{Register.}
We define $\gamma$ as an attribute tuple of smart contracts consisting of three elements ($x$, $y$, $z$). 
Among the attributes, $\gamma.x$ $\in$ \{0, 1\} is a flag to determine whether the contract is registered, $\gamma.y$ denotes the contract identifier of a contract and $\gamma.z$ means the address of the contract. 
To simplify the presentation, we assume that there is one requestor contract $C_r$ and one data owner contract $C_o$.
To utilize our protocol, $C_o$ is required to submit a registration request to $C_{DFC}$.
Once the request is received, $C_{DFC}$ will generate a list associated with $C_o$, which can be indexed by other contracts, like $C_r$.

Initially, a data owner $C_o$ must register with $C_{DFC}$.
The registration is send as :
\begin{equation}
    \alpha_{reg} = (I_{reg}, C_o, A_o)
\end{equation}
where $\alpha_{req}$ represents the registration request messgage, $I_{reg} \in \{0,1\}$ represents the registration interaction identifier. $C_o$ is the contract identifier of $C_o$ in the protocol, $A_o$ is the address of $C_o$.
Upon receiving this request, $C_{DFC}$ updates the tuple for $C_o$:
\begin{equation}
    \gamma_o = (1, C_o, K(A_o))
\end{equation}
where $\gamma_o$ is the attribute tuple for $C_o$ and $K(\cdot)$ is a keccak256 cryptographic hash function.
Only registered data owners can be discovered by requestor contracts.

\indent \textit{Request.} 
First, the requestor $C_r$ creates a request with the operation $I$.req, the name of $C_o$ and its own address.
Then, $C_r$ sends this request ($I_{req}$, $C_o$, $K(A_r)$) to $C_{DFM}$, waiting for the data  feed.

\indent \textit{Forward.}
For better security and iteration of \texttt{DeFeed}, our protocol has two pivotal smart contracts - a proxy contract ($C_{DFM}$) and a logic contract ($C_{DFC}$) that help to separate the data owner and requestor contracts.
The $C_{DFM}$ contract is purely a proxy contract and does not have any logical functions.
When it receives a request from $C_r$, it will check that the requestor contract $C_r$ has the necessary permissions to send the request. 
Once this check is passed, it will forward this request as ($I_{for}$, $C_o$, $K(A_r)$) to the $C_{DFC}$ contract.

\indent \textit{Query.}
Upon receiving ($I_{for}$, $C_o$, $K(A_r)$) from $C_{DFM}$, $C_{DFC}$ will search in $\gamma$ based on the passed username $C_o$.
If $\gamma.x$ is 0, it means that this name $Co$ is not registered yet and $C_{DFC}$ will send ($I_{resp}$, $K(A_r)$), $\beta$) to $C_r$, where $\beta$ is defined as a string and contains "The name doesn't exist.".
Otherwise, $C_{DFC}$ will send ($I_{query}$, $C_o$) to $C_o$ using the address of $C_o$ that is set in $\gamma$ before.

\indent \textit{Reply.}
Upon receiving the ($I_{query}$, $C_o$), $C_o$ will verify the sender's address. 
If the sender's address matches $C_{DFC}$, $C_o$ will execute the \textit{reply(·)} operation. In this operation, $C_o$ will retrieve the requested data and package it into the ($I_{reply}$, $\kappa$) response, where $\kappa$ is the encrypted data.
$C_o$ will then send this response back to the $C_{DFC}$ contract. 
However, if the sender's address does not match $C_{DFC}$, the \textit{accessOnly(·)} function in the code logic of $C_o$ will deny the $I_{query}$, preventing unauthorized access to the data.

\indent \textit{Respond.}
After processing the ($I_{reply}$, $\kappa$) response, the $C_{DFC}$ contract will perform several validation checks. 
It will verify that the response comes from $C_o$.
If the response is valid, $C_{DFC}$ will package the data into a final ($I_{resp}$, $\kappa$) message and send it back to the original requestor contract $C_r$. 
Then $C_r$ can retrieve and use the data that was requested from $C_o$. 
The $C_{DFC}$ contract also records information about the request and response in its internal logs or state for auditing and monitoring purposes.

\begin{figure}[!t]\small
\centering
\includegraphics[width=5.5in]{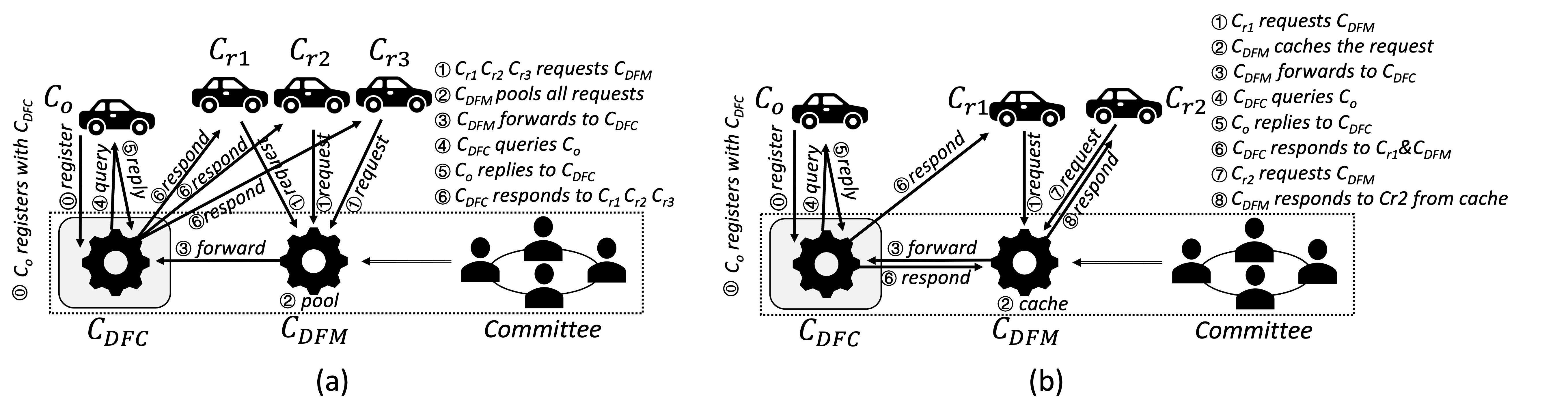}
\caption{ The protocol for Data Feed. (a) Pool  process, and (b) Cache process}
\label{fig:pool_cache}
\end{figure}

Next, we present the protocol with different functionalities for data feed and discuss how the four key components in the set $C$ work in various functionalities. 
We start by introducing the protocol with pool shown in Fig.~\ref{fig:pool_cache}(a), where $C_{DFM}$ can pool multiple requests and only send one request through function \textit{forward(·)}. 
We then introduce the protocol with cache in Fig.~\ref{fig:pool_cache}(b), where $C_{DFM}$ can cache any successful results from the previous contracts' request by the function \textit{respond(·)} in $C_{DFC}$.
When the result of another user request is found in $C_{DFM}$'s cache, the data can be obtained directly from the cache. 
Finally, we will respectively introduce the method that we use to realize the iteration of our protocol in the future and how users can subscribe to a target contract in the protocol.

\paragraph{Pool}
\label{subsec:pool}
To enhance the efficiency of data feed between different CAVs while reducing gas costs, we design the pool function that pools the requests with the same objective.
As outlined in Section \ref{subsec:datafeed}, the data feed with pool function also has the same four key components in the set $C$, but the functionality is kind of different in detail. 
In this scenario, we define a new set of the requestor contracts, $CR = \{C_{r1}, C_{r2}, ···, C_{rn}\}$ and redefine the attribute tuple $\gamma$ to include an additional element:$(\gamma.x, \gamma.y, \gamma.z, \gamma.t)$,
where $\gamma.x \in \{0,1\}$ indicates registration status, $\gamma.y$ is the registered username, $\gamma.z$ is the hashed address of the contract, and $\gamma.t$ specifies a transaction delay in seconds.
Initially, we set $\gamma.z$ to 12 seconds as the transaction delay, which is the time of one block submitted to the blockchain.
For clarity, we narrow down the number of request contracts to three, e.g., $CR'$ = $\{C_{r1}, C_{r2}, C_{r3}\}$.
The pool function operates as follows.

\textit{Register.}
In this phase, $C_o$ submits a registration request to $C_{DFC}$.
Upon receiving the request, $C_{DFC}$ updates $\gamma$ for $C_o$ as follows:
\begin{equation}
    \gamma_{o} = (1, C_o, K(A_o), -1)
\end{equation}
where $\gamma.t = -1$ indicates infinite waiting time, ensuring flexibility for initial registration.

\textit{Transmission.}
The transmission phase begins with a request from the first requestor, $C_{r1}$:
\begin{equation}
    \alpha_{req1} = (I_{req}, C_o, K(A_{r1}))
\end{equation}
Upon receiving this request, $C_{DFM}$ updates $\gamma.t$ for $C_o$ to 3 blocks, creating a time window of three blocks for other requests targeting $C_o$.
During this time, other requestors, $C_{r2}$ and $C_{r3}$, send their requests:
\begin{equation}
    \alpha_{req2} = (I_{req}, C_o, K(A_{r2}))
\end{equation}
\begin{equation}
    \alpha_{req3} = (I_{req}, C_o, K(A_{r3}))
\end{equation}
If these requests arrive within the time window, $C_{DFM}$ aggregates them into a single request:
\begin{equation}
    \alpha_{for} = (I_{for}, C_o, K(A_{r1}, A_{r2}, A_{r3}))
\end{equation}
This aggregated request is forwarded to $C_{DFC}$.

Upon receiving the forwarded request, $C_{DFC}$ searches for the name Co in its registry.
If $\gamma.reg = 0$, $C_{DFC}$ sends an error message back to $C_{r}$:
\begin{equation}
    \alpha_{resp} = (I_{resp}, K(A_{r}), \beta)
\end{equation}
Here, $\beta$ is a predefined string containing "The name doesn't exist."
In the case that $\gamma_x$ is not 0, this indicates that the name Co is already registered, $C_{DFC}$ proceeds to send the message ($I_{query}$, $C_o$) to $C_o$.
The address of $C_o$ used for sending this message is retrieved from $\gamma$.

Note that this transmission phase contains the forward, query, and reply phases.

\textit{Respond.}
Finally, $C_{DFC}$ processes the response and sends it back to each requestor:
\begin{equation}
    \alpha_{resp} = (I_{resp}, \kappa, K(A_r))
\end{equation}
The hashed address allows $C_{DFC}$ to identify and deliver the response to the corresponding requestors $C_{r1}$, $C_{r2}$ and $C_{r3}$.

Compared with the ordinary data feed mentioned, this way of processing requests can greatly shorten processing time and save gas cost.
By aggregating requests, the pool function enhances the efficiency and scalability of the data feed protocol while reducing the overall gas costs for participating contracts.

\paragraph{Cache}
\label{subsec:cache}
To further optimize gas consumption and improve the efficiency and scalability of the data feed protocol, we introduce a cache function based on the regular data feed.
This function leverages caching to reduce redundant interactions with the Data Owner Contract ($C_o$) by storing responses for subsequent reuse.

We redefine the attribute tuple $\gamma$ to include additional attributes for caching $(\gamma.x, \gamma.y, \gamma.z, \gamma.d)$, where $\gamma.x \in \{0,1\}$ indicates whether a cache exists for a given contract, $\gamma.y$ is the registered username of the contract, $\gamma.z$ is the hashed address of the contract and $\gamma.d$ stores the cached data.
The cache function uses the \textit{createCache(·)} operation to generate caches indexed by the username of the Data Owner Contract.
For clarity, we limit the number of requestors to two: $CR'$ = \{$C_{r1}$, $C_{r2}$\}.
Below, we describe the cache function process from the perspectives of $C_{r1}$ and $C_{r2}$.

\textit{Perspective CR1.}
When requestor $C_{r1}$ initiates a request:
\begin{equation}
    \alpha_{req} = (I_{req}, C_o, K(A_{r1}))
\end{equation}
the request follows the regular phases of the data feed protocol: \textbf{request}, \textbf{forward}, \textbf{query}, \textbf{reply}, and \textbf{respond}. During the respond phase, $C_{DFC}$ sends a response to both $C_{r1}$ and $C_{DFM}$:
\begin{equation}
    \alpha_{resp} = (I_{resp}, \kappa, K(A_{r1}))
\end{equation}
Upon receiving the response, $C_{DFM}$ executes the \textit{create Cache(·)} function to create a cache entry for $C_o$:
\begin{equation}
    \gamma_o =(1, C_o, K(A_o), \kappa)
\end{equation}
where $\gamma.x = 1$ indicates the cache is now active and $\gamma.d = \kappa$ stores the response data.

This caching ensures that future requests for $C_o$ can bypass redundant protocol phases.

\textit{Perspective CR2.}
When requestor $C_{r2}$ initiates a similar request:
\begin{equation}
    \alpha_{req} = (I_{req}, C_o, K(A_{r2}))
\end{equation}
$C_{DFM}$ checks whether a cache for $C_o$ exists by evaluating $\gamma.x$.
If $\gamma.x = 1$, $C_{DFM}$ directly sends  a cached response to $C_{r2}$:
\begin{equation}
    \alpha_{resp} = (I_{resp}, \kappa, K(A_{r2}))
\end{equation}
This skips the \textbf{forward}, \textbf{query}, \textbf{reply}, and part of the \textbf{respond} phases, significantly reducing gas costs and latency.
If $\gamma.x = 0$, the request follows all phases of the data feed protocol. Once the process completes, a cache is created for future use.

\paragraph{Update}
As blockchain technology evolves, smart contracts are becoming increasingly sophisticated and capable of executing more complex transactions.
However, like any software, smart contracts require periodic updates to enhance functionality, improve efficiency, and address security vulnerabilities. 
In \texttt{DeFeed}, we have implemented a robust update mechanism to ensure the protocol remains secure, efficient, and adaptable to changing requirements.

The update function in \texttt{DeFeed} serves multiple critical purposes. 
Primarily, it allows for security enhancements by enabling the patching of potential vulnerabilities identified through ongoing security audits. 
This proactive approach is crucial in maintaining the integrity of financial transactions handled by \texttt{DeFeed}. 
Furthermore, the update mechanism facilitates functionality improvements, allowing for the addition of new features and enhancement of existing ones. 
This not only improves user experience but also attracts more smart contracts to our protocol.

Adaptability is another key benefit of the update function. 
As the blockchain ecosystem evolves, this mechanism ensures \texttt{DeFeed} can adapt to new standards, regulations, or technological advancements. 
Additionally, updates can include optimizations that improve the protocol's efficiency, reducing gas costs and enhancing overall performance.

The update process in \texttt{DeFeed} is designed to be secure, efficient, and transparent. 
It begins with a multi-signature approval from the committee, ensuring that updates are consensual and aligned with the protocol's governance structure. Once approved, the Contract Data Feed Manager ($C_{DFM}$) executes the update function. This involves deactivating the previous Contract Data Feed Core ($C_{DFC}$) and replacing its address with the new one using the function $update(addr(newCDFC))$.

Following the update, extensive testing is conducted to ensure the new $C_{DFC}$ functions correctly and maintains backward compatibility where necessary. 
Users and stakeholders are then notified of the update, including any new features or changes in functionality.

The update mechanism in \texttt{DeFeed} is designed to have minimal disruption, ensuring a smooth transition to the new version with little to no downtime. 
Data integrity is maintained throughout the process, with all historical data and ongoing transactions preserved. 
This approach allows \texttt{DeFeed} to scale effectively, accommodating growing user bases and evolving blockchain ecosystems.

\begin{figure}[!t]\small
\centering
\includegraphics[width=3.5in]{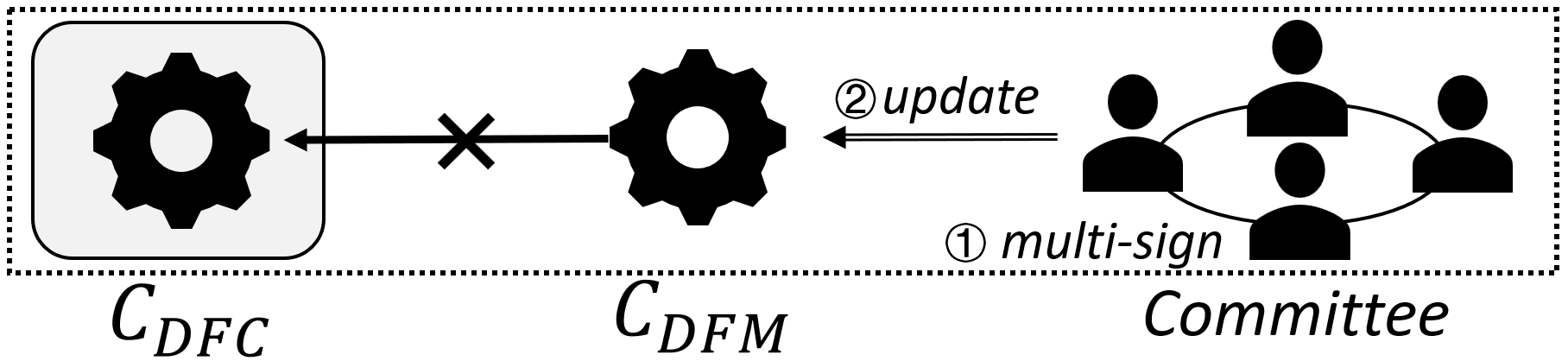}
\caption{ Update process}
\label{fig:update}
\end{figure}

Fig.~\ref{fig:update} illustrates the detailed update process, highlighting the role of the committee, $C_{DFM}$, and the transition from the old to the new $C_{DFC}$.
By implementing this comprehensive update mechanism, \texttt{DeFeed} ensures long-term viability, security, and adaptability in the dynamic landscape of blockchain technology. This approach not only addresses current needs but also positions the protocol to evolve alongside future technological advancements in the blockchain space.

\begin{figure}[!t]\small
\centering
\includegraphics[width=3.5in]{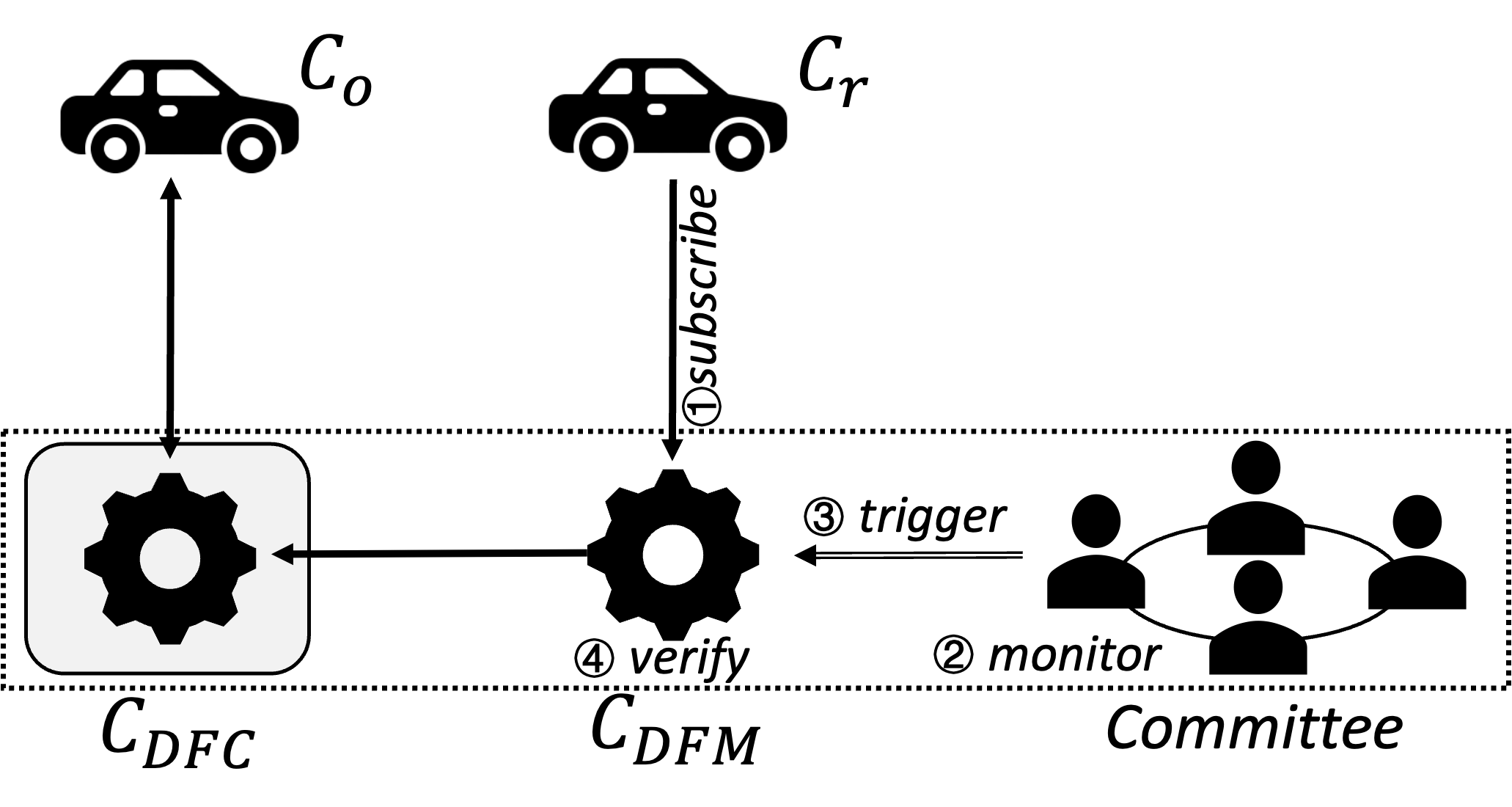}
\caption{ Subscribe process}
\label{fig:subscribe}
\end{figure}

\paragraph{Subscribe}
We now present a new function based on the regular data feed named subscribe. The subscribe process is shown in Fig.~\ref{fig:subscribe} and can be represented by the following equation:
\begin{equation}
\text{subscribe}(C_r, C_o, S) \rightarrow N(\Delta S)
\end{equation}

Where $C_r$ is the receiver contract (the one subscribing), $C_o$ is the observed contract (the one being subscribed to), $S$ represents the state of $C_o$, $\Delta$S represents any change in the status of $C_o$, and N($\Delta$S) is the notification containing details of the change.

The state $S$ of contract $C_o$ can be represented as a tuple:
\begin{equation}
S = (v_1, v_2, \ldots, v_n)
\end{equation}
where $v_i$ represents the $i$-th variable or property of the contract state.
When a change occurs, we can represent the new state $S'$ as:
\begin{equation}
S' = (v_1', v_2', \ldots, v_n')
\end{equation}
The change in state, $\Delta S$, can then be calculated as:
\begin{equation}
\Delta S = S' - S = (v_1' - v_1, v_2' - v_2, \ldots, v_n' - v_n)
\end{equation}
The notification function $N(\Delta S)$ can be defined as:
\begin{equation}
N(\Delta S) = \{(i, \Delta v_i) \mid \Delta v_i \neq 0, i \in [1,n]\}
\end{equation}
This function returns a set of pairs, where each pair consists of the index of the changed variable and its change value, for all non-zero changes.
In social media, we receive notifications when users update their status. Our \textit{subscribe(·)} function operates on a similar principle within the blockchain ecosystem. Under the supervision of the committee, other smart contracts, like $C_r$, can subscribe to the status of contract $C_o$ through the \textit{subscribe(·)} function.
The subscription process can be formalized as:
\begin{equation}
\text{Subscriptions}(C_o) = \{C_r \mid \text{subscribe}(C_r, C_o, S) \text{ has been called}\}
\end{equation}
This set represents all contracts that have subscribed to $C_o$.
Once there is any change ($\Delta S$) in the status of contract $C_o$, then for each $C_r \in \text{Subscriptions}(C_o)$, $C_r$ will immediately receive a notification $N(\Delta S)$, which contains the details of the change. 

In social media, we receive notifications when users update their status. Our \textit{subscribe(·)} function operates on a similar principle within the blockchain ecosystem. Under the supervision of the committee, other smart contracts, like $C_r$, can subscribe to the status of contract $C_o$ through the \textit{subscribe(·)} function.

Once there is any change ($\Delta$S) in the status of contract $C_o$, then $C_r$ will immediately receive a notification N($\Delta$S), which contains the details of the change. This makes it convenient for other smart contracts to know the changes in the status of the target contract anytime and anywhere, without having to check the contract status themselves.

This makes it convenient for other smart contracts to know the changes in the status of the target contract anytime and anywhere, without having to check the contract status themselves.
This subscribe mechanism creates a direct link between contracts, enabling efficient information flow and improving inter-contract communication. It allows smart contracts to stay informed about relevant changes in other contracts' states in real-time, enhancing overall system efficiency and responsiveness.

\subsection{Security analysis}
\label{sec:security-analysis}
In this section, we present the security analysis for \texttt{DeFeed}.
Our data feed protocol consists of a series of distinct smart contracts, enabling smart contracts to acquire the data they want from others utilizing the protocol at a minimal cost.\\

\noindent \textbf{Lemma 1.}
\textit{The data feed management contract $C_{DFM}$ exclusively accesses the data feed center contract $C_{DFC}$, which in turn solely accesses the data owner contract $C_o$.}\\

\noindent \textit{Proof.}
In Fig.~\ref{fig:regular}, the order of access begins with the requestor smart contract $C_r$, which is then passed on to the data feed management contract $C_{DFM}$. 
Next, the request sent by the requestor contract is forwarded to the data feed center contract $C_{DFC}$. 
Finally, the access request goes to the data owner contract $C_o$ and returns to its starting point, completing the cycle.

In order to ensure that other smart contracts do not access the contracts in the process arbitrarily, we design the data feed management contract, the data feed center contract, and the owner contract with \textit{accessOnly(·)} function.
Concretely, this function verifies whether the address of the accessing contract is the address of the specific contract.
If the verification is passed, the access proceeds normally. 
On the contrary, abnormal access will be rejected.
The verification process relies on blockchain technology, which ensures the accuracy and reliability of the verification performed on the contracts presented above.\\

\noindent \textbf{Lemma 2.}
\textit{As long as the owner smart contract is within access range, the requestor smart contract will be able to obtain the desired data.}\\

\noindent \textit{Proof.}
As we described in Section \ref{subsec:datafeed}, we highlight the significance of owner smart contracts during the registration phase and their critical role in the data feed process. 
In order to ensure that the requestor contract can obtain the desired data, we use function signatures at various stages of the data feed to achieve secure interaction between smart contracts.
Whenever the requestor contract calls a contract function via a transaction, the beginning of the transaction data contains the \textit{Keccak256} hash of the function signature, which tells the contract which function to execute.
Function signatures act as unique identifiers for functions in a smart contract, ensuring an exact match to the correct function when it is called. 
This prevents unexpected behavior due to incorrect function names or parameter types and allows contract developers to safely call functions from other contracts without being aware of the exact implementation details.\\

\noindent \textbf{Lemma 3.}
\textit{The integrity and security of the data feed management contract $C_{DFM}$ are maintained by the committee if the committee adheres to a threshold $t$ that $t < n/2$, where $n$ is the total number of committee members.}\\

\noindent \textit{Proof.}
The committee plays a crucial role in managing the $C_{DFM}$ contract, which acts as a central hub for coordinating interactions between the various components of the \texttt{DeFeed} protocol. 
Any modifications to the $C_{DFM}$ contract, including the updating of the $C_{DFC}$ address, are subject to a multi-signature approval process by the committee.
This process leverages a (t, n)-secret sharing scheme where $t < n/2$ ensures that no single entity, not even a compromised committee member, can unilaterally alter the contract. 
This ensures that the core components of the protocol remain secure and under the control of the trusted committee.

The committee also handles the registration and deregistration of data owner contracts ($C_o$) in the $C_{DFC}$ contract. 
By implementing a rigorous vetting process, the committee can significantly reduce the risk of including unauthorized or malicious contracts, which is essential for the protocol's resilience against adversarial attacks.
Furthermore, the committee actively monitors the \texttt{DeFeed} protocol's activities and performance. 
It possesses the authority to make data-driven adjustments to the protocol's parameters or governance rules. 
This committee-based adaptive governance is crucial for maintaining the protocol's security, efficiency, and liveness.

\section{Implementation and Evaluation}
\label{sec:imp}
In this section, we implement our protocol mentioned in Section \ref{sec:defeed} and evaluate each protocol's performance in detail.

\subsection{Implementation}
We program four key components and implement them with function pool, cache, update, and subscribe in Solidity, the contract-oriented advanced programming language for writing smart contracts. 
We test all of them in turn over Ethereum's official test network, Sepolia \footnote{\url{https://sepolia.etherscan.io/}}.

Next, we implement our basic protocol for realizing data feed. This protocol contains a smart contract $C_o$ that owns the original data, a requestor smart contract $C_r$ that wants to request the data, a data feed management smart contract $C_{DFM}$ for forwarding \textit{request} sending from $C_r$ and a data feed center smart contract $C_{DFC}$ for process the query information from $C_o$ and the forward information from $C_{DFM}$.
In order to achieve the Subscribe function, we design and implement several functions in different contracts respectively.  
In smart contract $C_r$, we design and implement the \textit{request(·)} function and the \textit{subscribe(·)} function, respectively. 
The function \textit{request(·)} aims to enable contract $C_r$ to request contract $C_{DFM}$ with the target contract name.
The function \textit{subscribe(·)} allows contract $C_r$ to subscribe to specific contract information from $C_{DFM}$.
The function \textit{forward(·)} and \textit{getSubscribe(·)} are implemented in contract $C_{DFM}$.
The former forward $C_r$'s request to $C_{DFC}$, and the latter accepts subscriptions and synchronizes subscription information to $C_{DFC}$.
In the contract $C_{DFC}$, the function \textit{query(·)} and \textit{respond(·)} are designed and implemented.
One is used to request data from the data owner $C_o$, and the other is used to feed data to $C_r$.

\begin{table}[!ht]\small
\centering
\caption{Gas Costs in Gas and USD}
\begin{tabular}{@{}lcr@{\hskip 0.5in}lcr@{}}
\toprule
\textbf{Function} & \textbf{Gas} & \textbf{USD} & \textbf{Function} & \textbf{Gas} & \textbf{USD} \\
\midrule
Deploy $C_{DFM}$ & 874393 & \$6.12 & Deploy $C_{DFC}$ & 1427517 & \$9.90 \\
Request & 143781 & \$1.00 & Update & 33241 & \$0.23 \\
Subscribe & 50094 & \$0.35 &  &  &  \\
\cmidrule{1-6}
Cache(initial) & 221668 & \$1.55 & Cache(subsequent) & 60145 & \$0.42\\
\bottomrule
\end{tabular}
\label{tab:GasToUSD}
\end{table}

\subsection{Evaluation}
We assess the effectiveness of \texttt{DeFeed} by examining the deployment of all smart contracts on the Sepolia test network. 
The evaluation focuses on the incurred expenses, quantified through fees paid to blockchain miners for executing contract calls. 
Within the Ethereum ecosystem, the primary metric for assessing costs is \textit{gas}, which encapsulates various factors influencing transaction execution, including computational complexity and storage requirements.
Our attention is directed toward the deployment cost of the protocol and the ensuing gas consumption during its operation. 
We acknowledge the importance of these aspects in the context of our business and academic pursuits and, hence, strive to optimize them for optimal performance.

\noindent \textbf{Approach to evaluation.}
As done in recent work, our evaluation focuses on measuring the gas consumption of protocols mentioned in Section \ref{sec:defeed}.
In TABLE \ref{tab:GasToUSD}, we present a comprehensive overview of the primary functions within the programmed smart contracts that engage with protocol participants at various stages of the protocol. 
Additionally, we outline the associated costs of these functions in terms of Gas and USD. 
To compute the costs in USD, we utilize the formula $cost(USD) = cost(Gas) * GasToEther * EtherToUSD$, leveraging the median of $GasToEther$ and $EtherToUSD$ from data of price as documented in Etherscan \footnote{\url{https://etherscan.io/}}. 
Since the historical price of ETH is volatile, we use the median of historical prices to calculate costs.
Specifically, $GasToEther$ was determined to be $2.42 * 10^{-8}$ Ether/Gas, while $EtherToUSD$ was established as 289.42 USD/Ether. 
By applying these values, we accurately compute the costs of the functions in USD based on their respective Gas costs.

\noindent \textbf{Gas cost.}
As demonstrated in Table \ref{tab:GasToUSD}, the deployment of the two core components of our protocol incurs a little high initial gas costs, 0.87M gas (\$6.12) for $C_{DFM}$ and 1.42M gas (\$9.9) for $C_{DFC}$.
It is critical to note that these expenses are one-time expenditures.
Subsequent uses of the protocol do not require additional gas for these actions, thereby eliminating further deployment costs.

The operational costs of the protocol also warrant discussion. 
Each smart contract interaction, referred to as a "Request", requires approximately 143,781 gas (\$1.00). For contracts subscribing to a target contract, the cost is significantly lower, at only 50,094 gas (\$0.35). This subscription mechanism reduces the need for repeated synchronization costs with the target contract, enhancing the protocol’s efficiency. Additionally, updating the core contract, $C_{DFC}$, is relatively inexpensive, requiring only 33,241 gas (\$0.23). These features collectively highlight the cost-effectiveness of maintaining the protocol infrastructure.

The Cache function within the protocol introduces further opportunities for gas optimization. While the initial request utilizing the Cache function consumes 221,668 gas (\$1.55), reflecting the additional computational resources required for caching, subsequent accesses benefit from a substantial reduction in cost, requiring only 60,145 gas (\$0.42). This marked decrease underscores the utility of the Cache function in minimizing long-term operational costs, making it an efficient option for repeated data retrievals within the protocol.

\begin{figure}[!ht]\small
\centering
\includegraphics[width=5.5in]{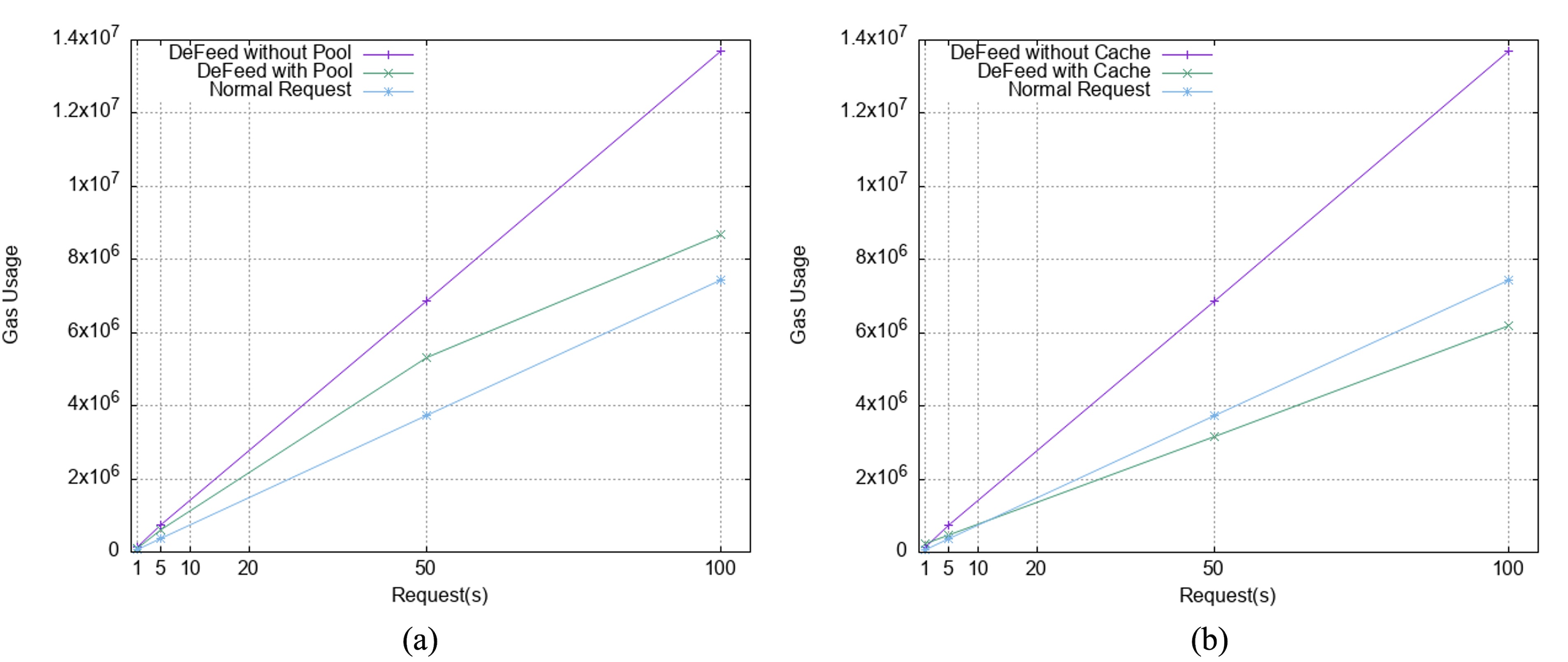}
\caption{Gas usage of DeFeed. (a) DeFeed with Pool, and (b) DeFeed with Cache}
\label{fig:poolGasUsage}
\end{figure}

\noindent \textbf{Gas saving.}
We compare the gas cost of \texttt{DeFeed} with and without the pool and cache mechanisms, which are introduced in Section \ref{subsec:pool} and Section \ref{subsec:cache}, respectively. We design six sets of controlled experiments, each deploying 1, 5, 10, 20, 50, and 100 requesting contracts. These smart contracts send requests to the same data owner contract.

As illustrated in Fig.~\ref{fig:poolGasUsage}(a), the implementation of the pool mechanism in \texttt{DeFeed} significantly reduces gas consumption. The graph clearly shows three distinct lines: "DeFeed without Pool" (purple), "DeFeed with Pool" (green), and "Normal Request" (blue).
The "Normal Request" mechanism represents the state-of-the-art method, embodying the basic interaction pattern between two contracts. While it shows the lowest gas consumption across all scenarios, with initial usage around 75,000 gas units for a single request, it lacks essential security considerations. This absence of security measures can expose users to vulnerabilities, making it less suitable for applications where trust and safety are paramount.

In contrast, the \texttt{DeFeed} mechanisms are designed with security in mind. Initially, both "DeFeed without Pool" and "DeFeed with Pool" start with similar gas usage for a single request, approximately 149,524 gas units. However, as the number of requests increases, the differences in gas consumption become more pronounced. At 20 requests, the "DeFeed with Pool" mechanism already shows a noticeable advantage, using about 20\% less gas than "DeFeed without Pool." By the time we reach 50 requests, the gap widens further, with "DeFeed with Pool" consuming around 5.33 million gas units compared to 6.87 million for "DeFeed without Pool," reflecting a reduction of nearly 22\%.

The most striking contrast occurs at 100 requests, where "DeFeed without Pool" consumes approximately 13.7 million gas units, while "DeFeed with Pool" requires only about 8.68 million gas units. This represents a remarkable reduction of around 36\%. The graph clearly illustrates that the pool function's efficiency increases with the number of requests, demonstrating how the benefits of this mechanism scale with higher transaction volumes.

Fig.~\ref{fig:poolGasUsage}(b) introduces the cache mechanism comparison. The graph shows three distinct lines: "DeFeed without Cache" (purple), "DeFeed with Cache" (green), and "Normal Request" (blue). Similar to the pool mechanism, the "Normal Request" method remains the most gas-efficient in terms of raw consumption, but it lacks security, making it unsuitable for secure applications.

For the cache mechanism, "DeFeed without Cache" starts with a gas usage of approximately 149,524 gas units for a single request, while "DeFeed with Cache" starts slightly higher, around 221,668 gas units. As the number of requests increases, the efficiency of the cache mechanism becomes evident. At 50 requests, "DeFeed with Cache" uses approximately 3.17 million gas units compared to 6.87 million for "DeFeed without Cache," indicating a reduction of about 53.8\%.

At 100 requests, the efficiency gains are even more pronounced. "DeFeed without Cache" consumes approximately 13.7 million gas units, while "DeFeed with Cache" requires only about 6.18 million gas units, resulting in a substantial reduction of around 54.9\%. This demonstrates that the cache mechanism significantly optimizes resource usage at higher volumes of transactions.

In summary, while the "Normal Request" method offers lower gas consumption, it does so at the cost of security, making it less viable for many applications. Our \texttt{DeFeed} protocol, particularly with the pool and cache functions, exhibits substantial efficiency gains, reducing gas usage by significant margins across different levels of request volume compared to "DeFeed without Pool" and "DeFeed without Cache." These performance improvements, clearly visualized in Fig.~\ref{fig:poolGasUsage}, indicate that the pool and cache mechanisms in \texttt{DeFeed} significantly optimize resource usage while maintaining essential security considerations, making them crucial features for gas optimization in the \texttt{DeFeed} protocol.

\begin{figure}[!ht]\small
\centering
\includegraphics[width=5.5in]{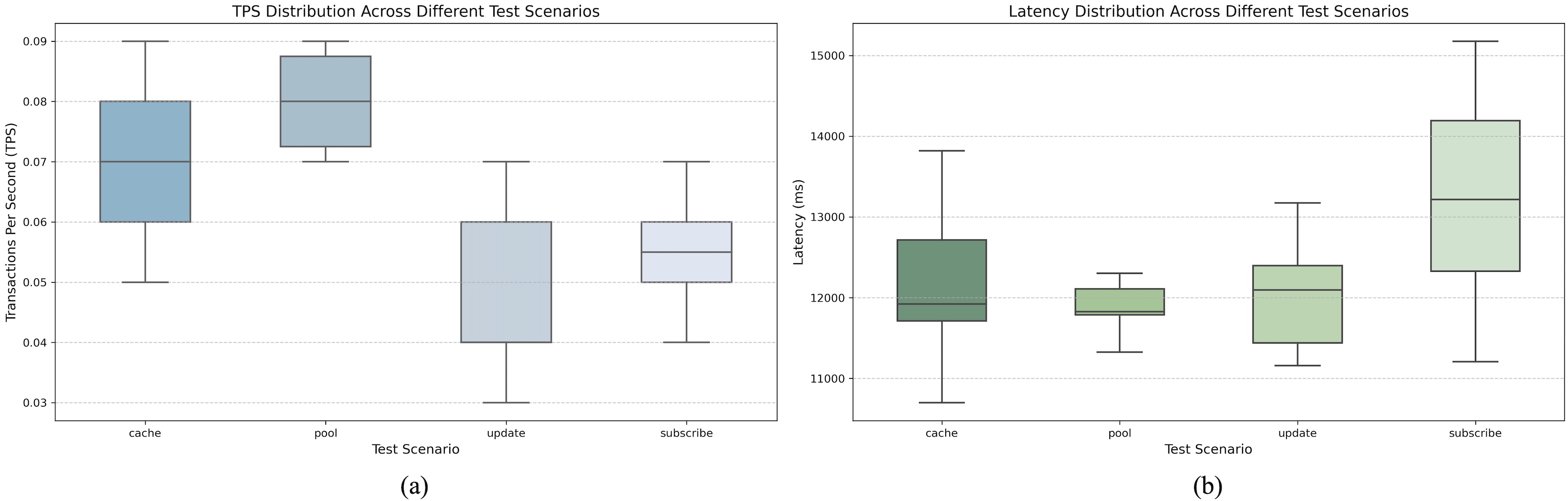}
\caption{Performance Metrics of  Different DeFeed Mechanisms. (a) Throughput (TPS) distribution across test scenarios, and (b) Latency distribution across test scenarios.}
\label{fig:tps-latency}
\end{figure}

\noindent \textbf{Throughput Testing.}
To evaluate the throughput of our \texttt{DeFeed} protocol, we conduct a series of performance tests using custom scripts. 
This script simulates multiple transactions and measures the system's ability to process these transactions over time. 
Our testing methodology focuses on key performance indicators including transactions per second (TPS), average latency, and gas consumption.

We implement the test script using Web3.js to interact with our smart contracts deployed on the Sepolia testnet. 
The script begins by initializing the connection to the Sepolia network and loading the necessary smart contract ABIs. It then sets a predetermined number of transactions to be executed. Before starting the transactions, the script records the start time of the test. The core of the script involves executing the specified number of transactions, each calling the \textit{request(·)} function of the $C_r$ contract.
Once all transactions are completed, the script records the end time and calculates the total duration of the test. The throughput is computed by dividing the number of successful transactions by the total duration. Additionally, the script calculates average latency and gas consumption for the transactions. To gain insights into network behavior, the script also analyzes block confirmation times.

Our test results show that the \texttt{DeFeed} protocol achieved an average throughput of 0.07 TPS under normal network conditions. The average latency for each transaction was 13881.40 milliseconds, with an average gas consumption of 65532 units per transaction.

To provide a more comprehensive view of the protocol's performance, we also examined the block confirmation times. The average block confirmation time was 13.89 seconds, with a minimum of 12.00 seconds and a maximum of 36.00 seconds. These figures help us understand the variability in transaction processing times due to network conditions and block creation intervals on the Sepolia testnet.

The boxplots in Fig.~\ref{fig:tps-latency}(a) illustrate the throughput performance across four \texttt{DeFeed} mechanisms: cache, pool, update, and subscribe. The pool mechanism demonstrates the highest median TPS (0.08), with a narrow interquartile range (IQR: 0.07–0.09), indicating stable performance due to optimized batch processing. In contrast, the update mechanism exhibits significant variability (median: 0.04, IQR: 0.03–0.07), reflecting inherent inefficiencies in cross-contract state synchronization. 
The cache strategy shows moderate throughput (median: 0.07) but occasional outliers (e.g., 0.05 and 0.09), likely caused by sporadic network disturbances. 
The subscribe mechanism achieves the lowest median TPS (0.06), constrained by event-driven overheads and asynchronous callback delays. These results quantitatively validate that batch-oriented designs (pool) outperform event-triggered approaches (subscribe) in high-frequency transaction environments.
Fig.~\ref{fig:tps-latency}(b) presents the latency characteristics of the same mechanisms. The cache and pool strategies show tightly clustered latencies (median: 11,798 ms and 11,816 ms, respectively), with 95\% of transactions completing within 13,000 ms, demonstrating predictable performance. 
The update mechanism displays severe right-skewing (median: 12,094 ms; maximum: 15,176 ms), attributable to recursive contract calls and gas contention during state updates. 
The subscribe mechanism shows the highest variation and the highest average, suggesting that subscribe-related operations are more complex.

It is essential to highlight that these results are specific to the Sepolia testnet environment and are inherently influenced by the network's constraints. Testnets like Sepolia often have lower transaction processing capacity compared to Ethereum's mainnet or other production environments. The relatively low TPS observed during our tests reflects these limitations, including network congestion and throttled block production intervals, which are designed to simulate realistic conditions but may not fully represent mainnet performance. Consequently, the throughput and latency metrics reported here should not be interpreted as hard limits of the \texttt{DeFeed} protocol's capabilities but rather as an indication of its performance under controlled, testnet-specific conditions.

Despite the testnet-imposed constraints, these throughput tests demonstrate that the \texttt{DeFeed} protocol can efficiently handle a significant number of requests, making it suitable for applications requiring frequent data feed updates. The relatively low average latency suggests that the protocol is capable of providing near real-time data feeds, a crucial feature for many decentralized applications.

Future work could include more extensive testing under various network conditions, simulating higher concurrency levels, and comparing the performance with other existing oracle solutions to provide a more comprehensive evaluation of \texttt{DeFeed}'s capabilities.

\section{Conclusion}
\label{sec:conclusion}
In this paper, we proposed \texttt{DeFeed}, a novel and secure protocol for decentralized cross-contract data feed in Web 3.0, specifically designed to enhance the functionality and efficiency of Connected Autonomous Vehicles networks.
Our approach leverages blockchain technology and smart contracts to provide a robust and scalable solution for inter-contract communication and data exchange.

We utilized a series of distinct smart contracts to construct our \texttt{DeFeed} protocol, which includes two central smart contracts, the data feed management smart contract ($C_{DFM}$) and the data feed center smart contract ($C_{DFC}$).
First, we introduced a regular data feed process that allows one smart contract to obtain data from another smart contract.
Recognizing the inefficiency of handling single requests at a time, we implemented the \texttt{Pool} function to enable simultaneous handling of multiple requests, thereby improving overall efficiency.
To further optimize resource utilization, we introduced the \texttt{Cache} function, which creates a new cache within $C_{DFM}$.
This function stores data obtained from target contracts, allowing subsequent requests for the same data to be fulfilled directly from the cache, thus reducing gas costs and enhancing performance. Additionally, we designed and implemented the \texttt{Update} function and \texttt{Subscribe} function. The \texttt{Update} function allows our protocol to replace the old data center contract $C_{DFC}$ with a new one, facilitating contract updates. The \texttt{Subscribe} function enables real-time information changes to be sent from the target contract to the requestor contract, allowing one contract to subscribe to another contract’s updates dynamically.

We implemented \texttt{DeFeed} over the Ethereum official test network, Sepolia. The results demonstrated that \texttt{DeFeed} is economical, convenient to implement, and efficient for facilitating communication and data feed between smart contracts. 
This makes \texttt{DeFeed} a ready-to-be-used protocol for CAV networks within the Web 3.0 ecosystem.
Our contributions to the field of autonomous and adaptive systems are significant. We highlighted the role of blockchain technology and smart contracts in achieving secure, decentralized, and efficient communication between entities in autonomous systems. By addressing the limitations of existing oracle-based solutions and focusing on inter-contract data exchange, we provided a practical framework for enhancing the adaptability and interconnectedness of decentralized applications.

The design and implementation of the \texttt{DeFeed} protocol showcase the potential of blockchain technology in decentralized autonomous systems and set a new standard for future research and development. \texttt{DeFeed}'s innovative features, including the Pool function for processing multiple requests simultaneously, the Cache function for optimizing resource utilization, and the Update and Subscribe functions for real-time data management, collectively enhance the adaptivity and efficiency of decentralized networks. The Pool function, in particular, addresses the challenge of handling multiple data requests efficiently, while the Cache mechanism significantly reduces gas costs and improves overall system performance. Furthermore, the Update and Subscribe functions ensure that the protocol remains flexible and responsive to changing data needs, crucial for systems like Connected Autonomous Vehicle networks that require real-time decision-making capabilities. These advancements extend beyond mere technical improvements, offering new paradigms for the design and operation of distributed systems. As the Web 3.0 ecosystem evolves, \texttt{DeFeed}'s comprehensive approach to inter-contract communication and data exchange positions it as a key enabler for more intelligent and autonomous decentralized applications, thereby shaping the future landscape of digital infrastructure.

While \texttt{DeFeed} shows promise, there are areas for future improvement. These include enhancing scalability for high-traffic scenarios, expanding cross-chain compatibility, and further optimizing energy efficiency. 
Additionally, as the regulatory landscape for blockchain technologies evolves, ensuring compliance across different jurisdictions will be an ongoing consideration. 
Future work will focus on addressing the identified limitations and exploring new use cases, further solidifying \texttt{DeFeed}'s position as a key protocol in the evolving landscape of decentralized technologies.

\bibliographystyle{ACM-Reference-Format}
\bibliography{DeFeed}


\begin{thebibliography}{52}


\ifx \showCODEN    \undefined \def \showCODEN     #1{\unskip}     \fi
\ifx \showDOI      \undefined \def \showDOI       #1{#1}\fi
\ifx \showISBNx    \undefined \def \showISBNx     #1{\unskip}     \fi
\ifx \showISBNxiii \undefined \def \showISBNxiii  #1{\unskip}     \fi
\ifx \showISSN     \undefined \def \showISSN      #1{\unskip}     \fi
\ifx \showLCCN     \undefined \def \showLCCN      #1{\unskip}     \fi
\ifx \shownote     \undefined \def \shownote      #1{#1}          \fi
\ifx \showarticletitle \undefined \def \showarticletitle #1{#1}   \fi
\ifx \showURL      \undefined \def \showURL       {\relax}        \fi
\providecommand\bibfield[2]{#2}
\providecommand\bibinfo[2]{#2}
\providecommand\natexlab[1]{#1}
\providecommand\showeprint[2][]{arXiv:#2}

\bibitem[Al-Breiki et~al\mbox{.}(2020)]%
        {trustworthy}
\bibfield{author}{\bibinfo{person}{Hamda Al-Breiki}, \bibinfo{person}{Muhammad
  Habib~Ur Rehman}, \bibinfo{person}{Khaled Salah}, {and}
  \bibinfo{person}{Davor Svetinovic}.} \bibinfo{year}{2020}\natexlab{}.
\newblock \showarticletitle{Trustworthy blockchain oracles: review, comparison,
  and open research challenges}.
\newblock \bibinfo{journal}{\emph{IEEE access}}  \bibinfo{volume}{8}
  (\bibinfo{year}{2020}), \bibinfo{pages}{85675--85685}.
\newblock


\bibitem[Ante(2022)]%
        {nft}
\bibfield{author}{\bibinfo{person}{Lennart Ante}.}
  \bibinfo{year}{2022}\natexlab{}.
\newblock \showarticletitle{The non-fungible token (NFT) market and its
  relationship with Bitcoin and Ethereum}.
\newblock \bibinfo{journal}{\emph{FinTech}} \bibinfo{volume}{1},
  \bibinfo{number}{3} (\bibinfo{year}{2022}), \bibinfo{pages}{216--224}.
\newblock


\bibitem[Antonino et~al\mbox{.}(2022)]%
        {safeCreationAndUpgradeOfEthereumSmartContracts}
\bibfield{author}{\bibinfo{person}{Pedro Antonino}, \bibinfo{person}{Juliandson
  Ferreira}, \bibinfo{person}{Augusto Sampaio}, {and} \bibinfo{person}{AW
  Roscoe}.} \bibinfo{year}{2022}\natexlab{}.
\newblock \showarticletitle{Specification is Law: Safe Creation and Upgrade of
  Ethereum Smart Contracts}. In \bibinfo{booktitle}{\emph{International
  Conference on Software Engineering and Formal Methods}}. Springer,
  \bibinfo{pages}{227--243}.
\newblock


\bibitem[Baza et~al\mbox{.}(2019)]%
        {baza2019blockchainfirmware}
\bibfield{author}{\bibinfo{person}{Mohamed Baza}, \bibinfo{person}{Mahmoud
  Nabil}, \bibinfo{person}{Noureddine Lasla}, \bibinfo{person}{Kemal Fidan},
  \bibinfo{person}{Mohamed Mahmoud}, {and} \bibinfo{person}{Mohamed Abdallah}.}
  \bibinfo{year}{2019}\natexlab{}.
\newblock \showarticletitle{Blockchain-based firmware update scheme tailored
  for autonomous vehicles}. In \bibinfo{booktitle}{\emph{2019 IEEE Wireless
  Communications and Networking Conference (WCNC)}}. IEEE,
  \bibinfo{pages}{1--7}.
\newblock


\bibitem[Bodell~III et~al\mbox{.}(2023)]%
        {proxyhunting}
\bibfield{author}{\bibinfo{person}{William~E Bodell~III},
  \bibinfo{person}{Sajad Meisami}, {and} \bibinfo{person}{Yue Duan}.}
  \bibinfo{year}{2023}\natexlab{}.
\newblock \showarticletitle{Proxy hunting: Understanding and characterizing
  proxy-based upgradeable smart contracts in blockchains}. In
  \bibinfo{booktitle}{\emph{32nd USENIX Security Symposium (USENIX Security
  23)}}. \bibinfo{pages}{1829--1846}.
\newblock


\bibitem[Bumiller et~al\mbox{.}(2023)]%
        {bumiller2023understanding}
\bibfield{author}{\bibinfo{person}{Anne Bumiller},
  \bibinfo{person}{St{\'e}phanie Challita}, \bibinfo{person}{Benoit Combemale},
  \bibinfo{person}{Olivier Barais}, \bibinfo{person}{Nicolas Aillery}, {and}
  \bibinfo{person}{Gael Le~Lan}.} \bibinfo{year}{2023}\natexlab{}.
\newblock \showarticletitle{On understanding context modelling for adaptive
  authentication systems}.
\newblock \bibinfo{journal}{\emph{ACM Transactions on Autonomous and Adaptive
  Systems}} \bibinfo{volume}{18}, \bibinfo{number}{1} (\bibinfo{year}{2023}),
  \bibinfo{pages}{1--35}.
\newblock


\bibitem[Buterin et~al\mbox{.}(2014)]%
        {buterin2014next}
\bibfield{author}{\bibinfo{person}{Vitalik Buterin} {et~al\mbox{.}}}
  \bibinfo{year}{2014}\natexlab{}.
\newblock \showarticletitle{A next-generation smart contract and decentralized
  application platform}.
\newblock \bibinfo{journal}{\emph{white paper}} \bibinfo{volume}{3},
  \bibinfo{number}{37} (\bibinfo{year}{2014}), \bibinfo{pages}{2--1}.
\newblock


\bibitem[Chen et~al\mbox{.}(2017)]%
        {Under-optimizedSmartContractsDevourYourMoney}
\bibfield{author}{\bibinfo{person}{Ting Chen}, \bibinfo{person}{Xiaoqi Li},
  \bibinfo{person}{Xiapu Luo}, {and} \bibinfo{person}{Xiaosong Zhang}.}
  \bibinfo{year}{2017}\natexlab{}.
\newblock \showarticletitle{Under-optimized smart contracts devour your money}.
  In \bibinfo{booktitle}{\emph{2017 IEEE 24th International Conference on
  Software Analysis, Evolution and Reengineering (SANER)}}. IEEE,
  \bibinfo{pages}{442--446}.
\newblock


\bibitem[Choi et~al\mbox{.}(2021)]%
        {smartian}
\bibfield{author}{\bibinfo{person}{Jaeseung Choi}, \bibinfo{person}{Doyeon
  Kim}, \bibinfo{person}{Soomin Kim}, \bibinfo{person}{Gustavo Grieco},
  \bibinfo{person}{Alex Groce}, {and} \bibinfo{person}{Sang~Kil Cha}.}
  \bibinfo{year}{2021}\natexlab{}.
\newblock \showarticletitle{Smartian: Enhancing smart contract fuzzing with
  static and dynamic data-flow analyses}. In \bibinfo{booktitle}{\emph{2021
  36th IEEE/ACM International Conference on Automated Software Engineering
  (ASE)}}. IEEE, \bibinfo{pages}{227--239}.
\newblock


\bibitem[Dameron(2019)]%
        {technicalSpecification}
\bibfield{author}{\bibinfo{person}{M Dameron}.}
  \bibinfo{year}{2019}\natexlab{}.
\newblock \showarticletitle{Beigepaper: An Ethereum Technical Specification v.
  0.8. 5}.
\newblock \bibinfo{journal}{\emph{Online}} (\bibinfo{year}{2019}).
\newblock


\bibitem[Feng et~al\mbox{.}(2020)]%
        {feng2020magmonitor}
\bibfield{author}{\bibinfo{person}{Yimeng Feng}, \bibinfo{person}{Guoqiang
  Mao}, \bibinfo{person}{Bo Chen}, \bibinfo{person}{Changle Li},
  \bibinfo{person}{Yilong Hui}, \bibinfo{person}{Zhigang Xu}, {and}
  \bibinfo{person}{Junliang Chen}.} \bibinfo{year}{2020}\natexlab{}.
\newblock \showarticletitle{MagMonitor: Vehicle speed estimation and vehicle
  classification through a magnetic sensor}.
\newblock \bibinfo{journal}{\emph{IEEE Transactions on Intelligent
  Transportation Systems}} \bibinfo{volume}{23}, \bibinfo{number}{2}
  (\bibinfo{year}{2020}), \bibinfo{pages}{1311--1322}.
\newblock


\bibitem[Feng et~al\mbox{.}(2019)]%
        {precise-attack-synthesis}
\bibfield{author}{\bibinfo{person}{Yu Feng}, \bibinfo{person}{Emina Torlak},
  {and} \bibinfo{person}{Rastislav Bodik}.} \bibinfo{year}{2019}\natexlab{}.
\newblock \showarticletitle{Precise attack synthesis for smart contracts}.
\newblock \bibinfo{journal}{\emph{arXiv preprint arXiv:1902.06067}}
  (\bibinfo{year}{2019}).
\newblock


\bibitem[Ferreira~Torres et~al\mbox{.}(2022)]%
        {elysium}
\bibfield{author}{\bibinfo{person}{Christof Ferreira~Torres},
  \bibinfo{person}{Hugo Jonker}, {and} \bibinfo{person}{Radu State}.}
  \bibinfo{year}{2022}\natexlab{}.
\newblock \showarticletitle{Elysium: Context-aware bytecode-level patching to
  automatically heal vulnerable smart contracts}. In
  \bibinfo{booktitle}{\emph{Proceedings of the 25th International Symposium on
  Research in Attacks, Intrusions and Defenses}}. \bibinfo{pages}{115--128}.
\newblock


\bibitem[Fiege et~al\mbox{.}(1987)]%
        {zero-knowledge-proof}
\bibfield{author}{\bibinfo{person}{Uriel Fiege}, \bibinfo{person}{Amos Fiat},
  {and} \bibinfo{person}{Adi Shamir}.} \bibinfo{year}{1987}\natexlab{}.
\newblock \showarticletitle{Zero knowledge proofs of identity}. In
  \bibinfo{booktitle}{\emph{Proceedings of the Nineteenth Annual ACM Symposium
  on Theory of Computing}}. \bibinfo{pages}{210--217}.
\newblock


\bibitem[Frank et~al\mbox{.}(2020)]%
        {ethbmc}
\bibfield{author}{\bibinfo{person}{Joel Frank}, \bibinfo{person}{Cornelius
  Aschermann}, {and} \bibinfo{person}{Thorsten Holz}.}
  \bibinfo{year}{2020}\natexlab{}.
\newblock \showarticletitle{{\{ETHBMC}\}: A bounded model checker for smart
  contracts}. In \bibinfo{booktitle}{\emph{29th USENIX Security Symposium
  (USENIX Security 20)}}. \bibinfo{pages}{2757--2774}.
\newblock


\bibitem[Grech et~al\mbox{.}(2018)]%
        {grech2018madmax}
\bibfield{author}{\bibinfo{person}{Neville Grech}, \bibinfo{person}{Michael
  Kong}, \bibinfo{person}{Anton Jurisevic}, \bibinfo{person}{Lexi Brent},
  \bibinfo{person}{Bernhard Scholz}, {and} \bibinfo{person}{Yannis
  Smaragdakis}.} \bibinfo{year}{2018}\natexlab{}.
\newblock \showarticletitle{Madmax: Surviving out-of-gas conditions in ethereum
  smart contracts}.
\newblock \bibinfo{journal}{\emph{Proceedings of the ACM on Programming
  Languages}} \bibinfo{volume}{2}, \bibinfo{number}{OOPSLA}
  (\bibinfo{year}{2018}), \bibinfo{pages}{1--27}.
\newblock


\bibitem[Guarnizo and Szalachowski(2019)]%
        {pdfs}
\bibfield{author}{\bibinfo{person}{Juan Guarnizo} {and} \bibinfo{person}{Pawel
  Szalachowski}.} \bibinfo{year}{2019}\natexlab{}.
\newblock \showarticletitle{PDFS: practical data feed service for smart
  contracts}. In \bibinfo{booktitle}{\emph{Computer Security--ESORICS 2019:
  24th European Symposium on Research in Computer Security, Luxembourg,
  September 23--27, 2019, Proceedings, Part I 24}}. Springer,
  \bibinfo{pages}{767--789}.
\newblock


\bibitem[Jabbar et~al\mbox{.}(2021)]%
        {jabbar2021V2X}
\bibfield{author}{\bibinfo{person}{Rateb Jabbar}, \bibinfo{person}{Noora
  Fetais}, \bibinfo{person}{Mohamed Kharbeche}, \bibinfo{person}{Moez Krichen},
  \bibinfo{person}{Kamel Barkaoui}, {and} \bibinfo{person}{Mohammed Shinoy}.}
  \bibinfo{year}{2021}\natexlab{}.
\newblock \showarticletitle{Blockchain for the Internet of Vehicles: How to use
  blockchain to secure vehicle-to-everything (V2X) communication and payment?}
\newblock \bibinfo{journal}{\emph{IEEE Sensors Journal}} \bibinfo{volume}{21},
  \bibinfo{number}{14} (\bibinfo{year}{2021}), \bibinfo{pages}{15807--15823}.
\newblock


\bibitem[Jain et~al\mbox{.}(2021)]%
        {jain2021blockchainAVs}
\bibfield{author}{\bibinfo{person}{Saurabh Jain}, \bibinfo{person}{Neelu~Jyothi
  Ahuja}, \bibinfo{person}{P Srikanth}, \bibinfo{person}{Kishor~Vinayak
  Bhadane}, \bibinfo{person}{Bharathram Nagaiah}, \bibinfo{person}{Adarsh
  Kumar}, {and} \bibinfo{person}{Charalambos Konstantinou}.}
  \bibinfo{year}{2021}\natexlab{}.
\newblock \showarticletitle{Blockchain and autonomous vehicles: Recent advances
  and future directions}.
\newblock \bibinfo{journal}{\emph{IEEE Access}}  \bibinfo{volume}{9}
  (\bibinfo{year}{2021}), \bibinfo{pages}{130264--130328}.
\newblock


\bibitem[Jiang et~al\mbox{.}(2018b)]%
        {contractfuzzer}
\bibfield{author}{\bibinfo{person}{Bo Jiang}, \bibinfo{person}{Ye Liu}, {and}
  \bibinfo{person}{Wing~Kwong Chan}.} \bibinfo{year}{2018}\natexlab{b}.
\newblock \showarticletitle{Contractfuzzer: Fuzzing smart contracts for
  vulnerability detection}. In \bibinfo{booktitle}{\emph{Proceedings of the
  33rd ACM/IEEE International Conference on Automated Software Engineering}}.
  \bibinfo{pages}{259--269}.
\newblock


\bibitem[Jiang et~al\mbox{.}(2018a)]%
        {jiang2018blockchainiov}
\bibfield{author}{\bibinfo{person}{Tigang Jiang}, \bibinfo{person}{Hua Fang},
  {and} \bibinfo{person}{Honggang Wang}.} \bibinfo{year}{2018}\natexlab{a}.
\newblock \showarticletitle{Blockchain-based internet of vehicles: Distributed
  network architecture and performance analysis}.
\newblock \bibinfo{journal}{\emph{IEEE Internet of Things Journal}}
  \bibinfo{volume}{6}, \bibinfo{number}{3} (\bibinfo{year}{2018}),
  \bibinfo{pages}{4640--4649}.
\newblock


\bibitem[Kosba et~al\mbox{.}(2016)]%
        {hawk}
\bibfield{author}{\bibinfo{person}{Ahmed Kosba}, \bibinfo{person}{Andrew
  Miller}, \bibinfo{person}{Elaine Shi}, \bibinfo{person}{Zikai Wen}, {and}
  \bibinfo{person}{Charalampos Papamanthou}.} \bibinfo{year}{2016}\natexlab{}.
\newblock \showarticletitle{Hawk: The blockchain model of cryptography and
  privacy-preserving smart contracts}. In \bibinfo{booktitle}{\emph{2016 IEEE
  Symposium on Security and Privacy (SP)}}. IEEE, \bibinfo{pages}{839--858}.
\newblock


\bibitem[Li and Palanisamy(2020)]%
        {li2020eventwarden}
\bibfield{author}{\bibinfo{person}{Chao Li} {and} \bibinfo{person}{Balaji
  Palanisamy}.} \bibinfo{year}{2020}\natexlab{}.
\newblock \showarticletitle{Eventwarden: A decentralized event-driven proxy
  service for outsourcing arbitrary transactions in ethereum-like blockchains}.
  In \bibinfo{booktitle}{\emph{2020 IEEE International Conference on Web
  Services (ICWS)}}. IEEE, \bibinfo{pages}{9--16}.
\newblock


\bibitem[Li and Palanisamy(2021)]%
        {li2021silentdelivery}
\bibfield{author}{\bibinfo{person}{Chao Li} {and} \bibinfo{person}{Balaji
  Palanisamy}.} \bibinfo{year}{2021}\natexlab{}.
\newblock \showarticletitle{Silentdelivery: Practical timed-delivery of private
  information using smart contracts}.
\newblock \bibinfo{journal}{\emph{IEEE Transactions on Services Computing}}
  \bibinfo{volume}{15}, \bibinfo{number}{6} (\bibinfo{year}{2021}),
  \bibinfo{pages}{3528--3540}.
\newblock


\bibitem[Li and Palanisamy(2024)]%
        {li2024twatch}
\bibfield{author}{\bibinfo{person}{Chao Li} {and} \bibinfo{person}{Balaji
  Palanisamy}.} \bibinfo{year}{2024}\natexlab{}.
\newblock \showarticletitle{T-Watch: Towards Timed Execution of Private
  Transaction in Blockchains}.
\newblock \bibinfo{journal}{\emph{IEEE Transactions on Services Computing}}
  (\bibinfo{year}{2024}).
\newblock


\bibitem[Li et~al\mbox{.}(2023)]%
        {li2023howhard}
\bibfield{author}{\bibinfo{person}{Chao Li}, \bibinfo{person}{Balaji
  Palanisamy}, \bibinfo{person}{Runhua Xu}, \bibinfo{person}{Li Duan},
  \bibinfo{person}{Jiqiang Liu}, {and} \bibinfo{person}{Wei Wang}.}
  \bibinfo{year}{2023}\natexlab{}.
\newblock \showarticletitle{How hard is takeover in dpos blockchains?
  understanding the security of coin-based voting governance}. In
  \bibinfo{booktitle}{\emph{Proceedings of the 2023 ACM SIGSAC Conference on
  Computer and Communications Security}}. \bibinfo{pages}{150--164}.
\newblock


\bibitem[Li et~al\mbox{.}(2020)]%
        {li2020nf}
\bibfield{author}{\bibinfo{person}{Chao Li}, \bibinfo{person}{Balaji
  Palanisamy}, \bibinfo{person}{Runhua Xu}, \bibinfo{person}{Jian Wang}, {and}
  \bibinfo{person}{Jiqiang Liu}.} \bibinfo{year}{2020}\natexlab{}.
\newblock \showarticletitle{Nf-crowd: Nearly-free blockchain-based
  crowdsourcing}. In \bibinfo{booktitle}{\emph{2020 International Symposium on
  Reliable Distributed Systems (SRDS)}}. IEEE, \bibinfo{pages}{41--50}.
\newblock


\bibitem[Li et~al\mbox{.}(2024)]%
        {li2024game}
\bibfield{author}{\bibinfo{person}{Nianyu Li}, \bibinfo{person}{Mingyue Zhang},
  \bibinfo{person}{Jialong Li}, \bibinfo{person}{Sridhar Adepu},
  \bibinfo{person}{Eunsuk Kang}, {and} \bibinfo{person}{Zhi Jin}.}
  \bibinfo{year}{2024}\natexlab{}.
\newblock \showarticletitle{A Game-Theoretical Self-Adaptation Framework for
  Securing Software-Intensive Systems}.
\newblock \bibinfo{journal}{\emph{ACM Transactions on Autonomous and Adaptive
  Systems}} \bibinfo{volume}{19}, \bibinfo{number}{2} (\bibinfo{year}{2024}),
  \bibinfo{pages}{1--49}.
\newblock


\bibitem[Liu et~al\mbox{.}(2024)]%
        {liu2024consortium}
\bibfield{author}{\bibinfo{person}{Bowen Liu}, \bibinfo{person}{Hao Tian},
  \bibinfo{person}{Zhijie Shen}, \bibinfo{person}{Yueyue Xu}, {and}
  \bibinfo{person}{Wanchun Dou}.} \bibinfo{year}{2024}\natexlab{}.
\newblock \showarticletitle{A Consortium Blockchain-Based Edge Task Offloading
  Method for Connected Autonomous Vehicles}.
\newblock \bibinfo{journal}{\emph{ACM Transactions on Autonomous and Adaptive
  Systems}} (\bibinfo{year}{2024}).
\newblock


\bibitem[Lo et~al\mbox{.}(2020)]%
        {reliability}
\bibfield{author}{\bibinfo{person}{Sin~Kuang Lo}, \bibinfo{person}{Xiwei Xu},
  \bibinfo{person}{Mark Staples}, {and} \bibinfo{person}{Lina Yao}.}
  \bibinfo{year}{2020}\natexlab{}.
\newblock \showarticletitle{Reliability analysis for blockchain oracles}.
\newblock \bibinfo{journal}{\emph{Computers \& Electrical Engineering}}
  \bibinfo{volume}{83} (\bibinfo{year}{2020}), \bibinfo{pages}{106582}.
\newblock


\bibitem[Luu et~al\mbox{.}(2016)]%
        {luu2016making}
\bibfield{author}{\bibinfo{person}{Loi Luu}, \bibinfo{person}{Duc-Hiep Chu},
  \bibinfo{person}{Hrishi Olickel}, \bibinfo{person}{Prateek Saxena}, {and}
  \bibinfo{person}{Aquinas Hobor}.} \bibinfo{year}{2016}\natexlab{}.
\newblock \showarticletitle{Making smart contracts smarter}. In
  \bibinfo{booktitle}{\emph{Proceedings of the 2016 ACM SIGSAC conference on
  computer and communications security}}. \bibinfo{pages}{254--269}.
\newblock


\bibitem[Megha et~al\mbox{.}(2020)]%
        {megha2020survey}
\bibfield{author}{\bibinfo{person}{Swati Megha}, \bibinfo{person}{Hamza Salem},
  \bibinfo{person}{Enes Ayan}, {and} \bibinfo{person}{Manuel Mazzara}.}
  \bibinfo{year}{2020}\natexlab{}.
\newblock \showarticletitle{A survey of blockchain solutions for autonomous
  vehicles ecosystems}. In \bibinfo{booktitle}{\emph{Journal of Physics:
  Conference Series}}, Vol.~\bibinfo{volume}{1694}. IOP Publishing,
  \bibinfo{pages}{012024}.
\newblock


\bibitem[Mingxiao et~al\mbox{.}(2017)]%
        {consensus-algorithm-review}
\bibfield{author}{\bibinfo{person}{Du Mingxiao}, \bibinfo{person}{Ma Xiaofeng},
  \bibinfo{person}{Zhang Zhe}, \bibinfo{person}{Wang Xiangwei}, {and}
  \bibinfo{person}{Chen Qijun}.} \bibinfo{year}{2017}\natexlab{}.
\newblock \showarticletitle{A review on consensus algorithm of blockchain}. In
  \bibinfo{booktitle}{\emph{2017 IEEE International Conference on Systems, Man,
  and Cybernetics (SMC)}}. IEEE, \bibinfo{pages}{2567--2572}.
\newblock


\bibitem[Nakamoto(2008)]%
        {nakamoto2008bitcoin}
\bibfield{author}{\bibinfo{person}{Satoshi Nakamoto}.}
  \bibinfo{year}{2008}\natexlab{}.
\newblock \showarticletitle{Bitcoin: A peer-to-peer electronic cash system}.
\newblock \bibinfo{journal}{\emph{Decentralized business review}}
  (\bibinfo{year}{2008}).
\newblock


\bibitem[Nguyen et~al\mbox{.}(2020)]%
        {sfuzz}
\bibfield{author}{\bibinfo{person}{Tai~D Nguyen}, \bibinfo{person}{Long~H
  Pham}, \bibinfo{person}{Jun Sun}, \bibinfo{person}{Yun Lin}, {and}
  \bibinfo{person}{Quang~Tran Minh}.} \bibinfo{year}{2020}\natexlab{}.
\newblock \showarticletitle{sfuzz: An efficient adaptive fuzzer for solidity
  smart contracts}. In \bibinfo{booktitle}{\emph{Proceedings of the ACM/IEEE
  42nd International Conference on Software Engineering}}.
  \bibinfo{pages}{778--788}.
\newblock


\bibitem[Rathee et~al\mbox{.}(2019)]%
        {rathee2019blockchainframework}
\bibfield{author}{\bibinfo{person}{Geetanjali Rathee},
  \bibinfo{person}{Ashutosh Sharma}, \bibinfo{person}{Razi Iqbal},
  \bibinfo{person}{Moayad Aloqaily}, \bibinfo{person}{Naveen Jaglan}, {and}
  \bibinfo{person}{Rajiv Kumar}.} \bibinfo{year}{2019}\natexlab{}.
\newblock \showarticletitle{A blockchain framework for securing connected and
  autonomous vehicles}.
\newblock \bibinfo{journal}{\emph{Sensors}} \bibinfo{volume}{19},
  \bibinfo{number}{14} (\bibinfo{year}{2019}), \bibinfo{pages}{3165}.
\newblock


\bibitem[Ren et~al\mbox{.}(2021)]%
        {ren2021double}
\bibfield{author}{\bibinfo{person}{Wei Ren}, \bibinfo{person}{Xutao Wan}, {and}
  \bibinfo{person}{Pengcheng Gan}.} \bibinfo{year}{2021}\natexlab{}.
\newblock \showarticletitle{A double-blockchain solution for agricultural
  sampled data security in Internet of Things network}.
\newblock \bibinfo{journal}{\emph{Future Generation Computer Systems}}
  \bibinfo{volume}{117} (\bibinfo{year}{2021}), \bibinfo{pages}{453--461}.
\newblock


\bibitem[Ritzdorf et~al\mbox{.}(2018)]%
        {signingtls}
\bibfield{author}{\bibinfo{person}{Hubert Ritzdorf}, \bibinfo{person}{Karl
  Wüst}, \bibinfo{person}{Arthur Gervais}, \bibinfo{person}{Guillaume Felley},
  {and} \bibinfo{person}{Srdjan Capkun}.} \bibinfo{year}{2018}\natexlab{}.
\newblock \showarticletitle{TLS-N: Non-repudiation over TLS Enabling Ubiquitous
  Content Signing}.
  \bibinfo{howpublished}{\url{http://dx.doi.org/10.14722/ndss.2018.23272}}. In
  \bibinfo{booktitle}{\emph{Network and Distributed Systems Security (NDSS)
  Symposium 2018}}. \bibinfo{pages}{18--21}.
\newblock


\bibitem[Shi et~al\mbox{.}(2023)]%
        {shi2023ress}
\bibfield{author}{\bibinfo{person}{Dongxian Shi}, \bibinfo{person}{Xiaoqing
  Wang}, \bibinfo{person}{Ming Xu}, \bibinfo{person}{Liang Kou}, {and}
  \bibinfo{person}{Hongbing Cheng}.} \bibinfo{year}{2023}\natexlab{}.
\newblock \showarticletitle{RESS: A reliable and effcient storage scheme for
  bitcoin blockchain based on raptor code}.
\newblock \bibinfo{journal}{\emph{Chinese Journal of Electronics}}
  \bibinfo{volume}{32}, \bibinfo{number}{3} (\bibinfo{year}{2023}),
  \bibinfo{pages}{577--586}.
\newblock


\bibitem[Singh et~al\mbox{.}(2020)]%
        {singh2020IoV}
\bibfield{author}{\bibinfo{person}{Pranav~Kumar Singh}, \bibinfo{person}{Roshan
  Singh}, \bibinfo{person}{Sunit~Kumar Nandi}, \bibinfo{person}{Kayhan~Zrar
  Ghafoor}, \bibinfo{person}{Danda~B Rawat}, {and} \bibinfo{person}{Sukumar
  Nandi}.} \bibinfo{year}{2020}\natexlab{}.
\newblock \showarticletitle{Blockchain-based adaptive trust management in
  internet of vehicles using smart contract}.
\newblock \bibinfo{journal}{\emph{IEEE Transactions on Intelligent
  Transportation Systems}} \bibinfo{volume}{22}, \bibinfo{number}{6}
  (\bibinfo{year}{2020}), \bibinfo{pages}{3616--3630}.
\newblock


\bibitem[Su and Jiang(2023)]%
        {su2023hybrid}
\bibfield{author}{\bibinfo{person}{Jian Su} {and} \bibinfo{person}{Mengnan
  Jiang}.} \bibinfo{year}{2023}\natexlab{}.
\newblock \showarticletitle{A hybrid entropy and blockchain approach for
  network security defense in SDN-based IIoT}.
\newblock \bibinfo{journal}{\emph{Chinese Journal of Electronics}}
  \bibinfo{volume}{32}, \bibinfo{number}{3} (\bibinfo{year}{2023}),
  \bibinfo{pages}{531--541}.
\newblock


\bibitem[Szabo et~al\mbox{.}(1994)]%
        {szabo1994smart}
\bibfield{author}{\bibinfo{person}{Nick Szabo} {et~al\mbox{.}}}
  \bibinfo{year}{1994}\natexlab{}.
\newblock \bibinfo{title}{Smart contracts}.
\newblock
\newblock


\bibitem[Wang et~al\mbox{.}(2022)]%
        {privacy-data-feed}
\bibfield{author}{\bibinfo{person}{Hao Wang}, \bibinfo{person}{Zhe Liu},
  \bibinfo{person}{Chunpeng Ge}, \bibinfo{person}{Kouichi Sakurai}, {and}
  \bibinfo{person}{Chunhua Su}.} \bibinfo{year}{2022}\natexlab{}.
\newblock \showarticletitle{A privacy-preserving data feed scheme for smart
  contracts}.
\newblock \bibinfo{journal}{\emph{IEICE TRANSACTIONS on Information and
  Systems}} \bibinfo{volume}{105}, \bibinfo{number}{2} (\bibinfo{year}{2022}),
  \bibinfo{pages}{195--204}.
\newblock


\bibitem[Wang et~al\mbox{.}(2019)]%
        {wang2019blockchain}
\bibfield{author}{\bibinfo{person}{Shuai Wang}, \bibinfo{person}{Liwei Ouyang},
  \bibinfo{person}{Yong Yuan}, \bibinfo{person}{Xiaochun Ni},
  \bibinfo{person}{Xuan Han}, {and} \bibinfo{person}{Fei-Yue Wang}.}
  \bibinfo{year}{2019}\natexlab{}.
\newblock \showarticletitle{Blockchain-enabled smart contracts: architecture,
  applications, and future trends}.
\newblock \bibinfo{journal}{\emph{IEEE Transactions on Systems, Man, and
  Cybernetics: Systems}} \bibinfo{volume}{49}, \bibinfo{number}{11}
  (\bibinfo{year}{2019}), \bibinfo{pages}{2266--2277}.
\newblock


\bibitem[Wang et~al\mbox{.}(2023)]%
        {wang2023fastv2v}
\bibfield{author}{\bibinfo{person}{Yingsen Wang}, \bibinfo{person}{Leiming
  Yuan}, \bibinfo{person}{Weihan Jiao}, \bibinfo{person}{Yan Qiang},
  \bibinfo{person}{Juanjuan Zhao}, \bibinfo{person}{Qianqian Yang}, {and}
  \bibinfo{person}{Keqin Li}.} \bibinfo{year}{2023}\natexlab{}.
\newblock \showarticletitle{A fast and secured vehicle-to-vehicle energy
  trading based on blockchain consensus in the internet of electric vehicles}.
\newblock \bibinfo{journal}{\emph{IEEE Transactions on Vehicular Technology}}
  \bibinfo{volume}{72}, \bibinfo{number}{6} (\bibinfo{year}{2023}),
  \bibinfo{pages}{7827--7843}.
\newblock


\bibitem[Wood et~al\mbox{.}(2014)]%
        {wood2014ethereum}
\bibfield{author}{\bibinfo{person}{Gavin Wood} {et~al\mbox{.}}}
  \bibinfo{year}{2014}\natexlab{}.
\newblock \showarticletitle{Ethereum: A secure decentralised generalised
  transaction ledger}.
\newblock \bibinfo{journal}{\emph{Ethereum project yellow paper}}
  \bibinfo{volume}{151}, \bibinfo{number}{2014} (\bibinfo{year}{2014}),
  \bibinfo{pages}{1--32}.
\newblock


\bibitem[Xu et~al\mbox{.}(2020)]%
        {multimediaWorkflow}
\bibfield{author}{\bibinfo{person}{Xiaolong Xu}, \bibinfo{person}{Yi Chen},
  \bibinfo{person}{Yuan Yuan}, \bibinfo{person}{Tao Huang},
  \bibinfo{person}{Xuyun Zhang}, {and} \bibinfo{person}{Lianyong Qi}.}
  \bibinfo{year}{2020}\natexlab{}.
\newblock \showarticletitle{Blockchain-based cloudlet management for multimedia
  workflow in mobile cloud computing}.
\newblock \bibinfo{journal}{\emph{Multimedia Tools and Applications}}
  \bibinfo{volume}{79} (\bibinfo{year}{2020}), \bibinfo{pages}{9819--9844}.
\newblock


\bibitem[Xu et~al\mbox{.}(2022)]%
        {xu2022reputation}
\bibfield{author}{\bibinfo{person}{Xiaolong Xu}, \bibinfo{person}{Ji Gu},
  \bibinfo{person}{Hanzhi Yan}, \bibinfo{person}{Wentao Liu},
  \bibinfo{person}{Lianyong Qi}, {and} \bibinfo{person}{Xiaokang Zhou}.}
  \bibinfo{year}{2022}\natexlab{}.
\newblock \showarticletitle{Reputation-aware supplier assessment for
  blockchain-enabled supply chain in industry 4.0}.
\newblock \bibinfo{journal}{\emph{IEEE Transactions on Industrial Informatics}}
  \bibinfo{volume}{19}, \bibinfo{number}{4} (\bibinfo{year}{2022}),
  \bibinfo{pages}{5485--5494}.
\newblock


\bibitem[Xu et~al\mbox{.}(2021)]%
        {xu2021concurrent}
\bibfield{author}{\bibinfo{person}{Xiaolong Xu}, \bibinfo{person}{Dawei Zhu},
  \bibinfo{person}{Xiaoxian Yang}, \bibinfo{person}{Shuo Wang},
  \bibinfo{person}{Lianyong Qi}, {and} \bibinfo{person}{Wanchun Dou}.}
  \bibinfo{year}{2021}\natexlab{}.
\newblock \showarticletitle{Concurrent practical byzantine fault tolerance for
  integration of blockchain and supply chain}.
\newblock \bibinfo{journal}{\emph{ACM Transactions on Internet Technology
  (TOIT)}} \bibinfo{volume}{21}, \bibinfo{number}{1} (\bibinfo{year}{2021}),
  \bibinfo{pages}{1--17}.
\newblock


\bibitem[Xue et~al\mbox{.}(2022)]%
        {xue2022xfuzz}
\bibfield{author}{\bibinfo{person}{Yinxing Xue}, \bibinfo{person}{Jiaming Ye},
  \bibinfo{person}{Wei Zhang}, \bibinfo{person}{Jun Sun}, \bibinfo{person}{Lei
  Ma}, \bibinfo{person}{Haijun Wang}, {and} \bibinfo{person}{Jianjun Zhao}.}
  \bibinfo{year}{2022}\natexlab{}.
\newblock \showarticletitle{xfuzz: Machine learning guided cross-contract
  fuzzing}.
\newblock \bibinfo{journal}{\emph{IEEE Transactions on Dependable and Secure
  Computing}} (\bibinfo{year}{2022}).
\newblock


\bibitem[Yang et~al\mbox{.}(2023)]%
        {yang2023zero}
\bibfield{author}{\bibinfo{person}{Kunwei Yang}, \bibinfo{person}{Bo Yang},
  \bibinfo{person}{Tao Wang}, {and} \bibinfo{person}{Yanwei Zhou}.}
  \bibinfo{year}{2023}\natexlab{}.
\newblock \showarticletitle{Zero-cerd: a self-blindable anonymous
  authentication system based on blockchain}.
\newblock \bibinfo{journal}{\emph{Chinese Journal of Electronics}}
  \bibinfo{volume}{32}, \bibinfo{number}{3} (\bibinfo{year}{2023}),
  \bibinfo{pages}{587--596}.
\newblock


\bibitem[Zhang et~al\mbox{.}(2016)]%
        {town-crier}
\bibfield{author}{\bibinfo{person}{Fan Zhang}, \bibinfo{person}{Ethan
  Cecchetti}, \bibinfo{person}{Kyle Croman}, \bibinfo{person}{Ari Juels}, {and}
  \bibinfo{person}{Elaine Shi}.} \bibinfo{year}{2016}\natexlab{}.
\newblock \showarticletitle{Town crier: An authenticated data feed for smart
  contracts}. In \bibinfo{booktitle}{\emph{Proceedings of the 2016 ACM SIGSAC
  Conference on Computer and Communications Security}}.
  \bibinfo{pages}{270--282}.
\newblock


\end{thebibliography}

\end{document}